\documentclass{jfm}

\usepackage{graphicx}
\usepackage{newtxtext}
\usepackage{newtxmath}
\usepackage{natbib}
\usepackage{hyperref}
\hypersetup{
    colorlinks = true,
    urlcolor   = blue,
    citecolor  = black,
}

\usepackage[dvipsnames]{xcolor} %
\usepackage{amsmath,amssymb} 
\usepackage{subcaption}

\title{A static-core framework for the slump of an axisymmetric granular column}

\author{Igor Le Pivert-{}-Jolivet\aff{1}\corresp{\email{igor.le-pivert-jolivet@universite-paris-saclay.fr}}, Cyprien Morize\aff{1,2}\corresp{\email{cyprien.morize@universite-paris-saclay.fr}}, Alban Sauret\aff{3,4}, Philippe Gondret\aff{1}}

\affiliation{
\aff{1}Université Paris-Saclay, CNRS, Laboratoire FAST, F-91405 Orsay, France
\aff{2}Institut Franco-Argentin de Dynamique des Fluides pour l’Environnement, CNRS (IRL 2027),
Universidad de Buenos Aires, CONICET, Buenos Aires, Argentina
\aff{3}Department of Mechanical Engineering, University of Maryland, College Park, Maryland 20742, USA
\aff{4}Department of Chemical and Biomolecular Engineering, University of Maryland, College Park, Maryland 20742, USA
}

\providecommand{\ie}{i.e.\ }

\newcommand{\circletfill}{\ensuremath{\bullet}}
\newcommand{\squadfill}{\ensuremath{\blacksquare}}
\newcommand{\triangleplfill}{\ensuremath{\blacktriangleleft}}
\newcommand{\triangleprfill}{\ensuremath{\blacktriangleright}}

\usepackage{etoolbox}
\makeatletter
\patchcmd{\@maketitle}{(Received xx; revised xx; accepted xx)\hfill}{}{}{\PackageError{arxiv-preprint}{Could not remove placeholder dates}{}}
\let\absfooterflag\relax
\let\pagelimitfooter\relax
\makeatother

\begin{document}
\maketitle
\thispagestyle{plain}
\begin{abstract}
Gravity-driven failures of granular masses, such as landslides and rockfalls, are often modelled as the collapse of initially confined granular columns. We experimentally investigate the dry collapse of axisymmetric granular columns focusing on deposit geometry, mass partition, and energy dissipation over a wide range of aspect ratios $a=H_0/R_0$. The deposits exhibit three regimes: truncated cones at low $a$, cones at intermediate $a$, and granular-jet-like deposits at large $a$. In regimes I and II, image differencing shows that only an outer shell is mobilised, while a static inner core persists throughout the collapse. The interface delimiting this core has a nearly constant slope angle, $\theta_2 \simeq 37^\circ$. This observation provides a simple geometric framework that rationalises the transition between deposit shapes, the final height and runout scalings, and the partition between flowing and static mass. By tracking the barycentre of an elementary azimuthal slice, we estimate the potential-energy loss and infer a global effective friction coefficient, which increases with both $a$ and a macroscopic inertial number. Finally, comparisons with terrestrial and Martian landslide data indicate that the laboratory runout and dissipation trends provide a useful reference for {comparison with} natural granular failures.
\end{abstract}


\section{Introduction}

Gravity-driven failures of granular masses, such as landslides, rockfalls, and debris avalanches, are ubiquitous in natural environments and can generate severe hazards \citep{Delannay2017}. A central challenge for hazard assessment is predicting how far a given unstable mass will travel once mobilised, and understanding how its initial gravitational potential energy is ultimately converted into motion and dissipation during the event
\citep{DadeHuppert1998,Legros2002}. Systematic compilations of field data have revealed robust trends between runout, volume, and fall height for long-runout landslides on Earth and on Mars, as well as the diversity of failure modes and materials involved \citep{Legros2002,Hungr2014,Lajeunesse2006Mars}. Despite this diversity, a common feature of these events is that they involve dense granular media evolving under gravity, often in regimes where interstitial fluids and cohesion play only a secondary role.

Numerous studies have focused on the collapse of columns of grains onto a horizontal plane to isolate the basic ingredients that control such gravity-driven granular flows, as compiled in the recent review by \cite{Cabrera2025}, which includes a large open database of up to 500 studies. In this canonical configuration, a column of dry grains is suddenly released by removing a gate, and the material collapses and spreads until it forms a final deposit. This simple experiment has been performed in a variety of geometries, including quasi-two-dimensional slots and rectangular channels \citep{BalmforthKerswell2005,Lajeunesse2005Slumping,Staron2005,Lacaze2008,Sarlin2021PRE} and axisymmetric configurations based on initially cylindrical columns \citep{Lajeunesse2004Spreading,Lube2004,Thompson2007,Lacaze2009}. In axisymmetric set-ups, experiments have been carried out with full or half-cylindrical columns \citep[e.g.][]{Lajeunesse2005Slumping}, and these studies show that, within experimental uncertainty, the global dynamics of the collapse is essentially insensitive to this angular extent. Using a half cylinder, however, has the practical advantage that the free sidewall provides optical access to the interior of the column, allowing the internal structure and grain motion to be visualised directly. In all cases, the dynamics and morphology of the collapse are controlled primarily by the initial aspect ratio of the column, defined as $a = H_0/R_0$ for a column of initial height $H_0$ and radius $R_0$ (or $a = H_0/L_0$ in planar geometries with the initial length $L_0$). Short columns topple by avalanching of their flanks and produce truncated deposits, whereas tall columns undergo a more vigorous collapse and form extended conical deposits. The corresponding runout distances and deposit heights follow simple power-law scalings with $a$, which have been rationalised using free-fall or shallow-flow arguments \citep{Lajeunesse2004Spreading,Lube2004,BalmforthKerswell2005,Staron2005,Larrieu2006}.

Beyond these global scalings, several studies have explored how grain properties, geometric confinement, and additional physical effects modify the collapse. Numerical simulations have shown how grain shape, restitution, interparticle friction, {and basal friction} affect the internal flow structure, deposit geometry and runout, while leaving the aspect ratio as a primary control parameter \citep{StaronHinch2007,Lai2023,Li2024}. { \cite{Man2022} have also explored the non-trivial effect of a non-circular cross section, such as square, rectangular or triangular, on the deposition morphology and directional runout.} 
Other work has examined the effect of interstitial fluids or partial saturation on the collapse dynamics and deposit morphology, showing that pore pressure and fluid drag can enhance mobility in specific regimes \cite[\textit{e.g.},][]{Rondon2011,bougouin2019collapse}{, with a strong influence of the initial packing fraction of the column}. More recently,~\cite{Sharma2024} have demonstrated that interparticle cohesion, either capillary or polymer-based, can strongly alter the final shape and runout of collapsing granular columns, and proposed a unified framework based on a macroscopic cohesion number. In the present study, we restrict attention to cohesionless, dry grains on a rough substrate to focus on the roles of geometry, internal friction, and basal roughness.

A key feature of granular column collapses, already emphasised in the early experiments~\citep{Lajeunesse2004Spreading,Lajeunesse2005Slumping,Lube2004,Lube2007,Man2021}, {is that only a fraction of the granular column flows, while the remaining grains form a static interior.} In axisymmetric collapses of cylindrical columns, these studies consistently report the presence of a stagnant inner cone attached to the axis, surrounded by a flowing annulus that ultimately forms the peripheral part of the final deposit. Two-dimensional discrete-grain simulations have confirmed the existence of such stagnant cores and highlighted the nontrivial manner in which the flowing region thickens as the aspect ratio increases \citep{Staron2005,StaronHinch2007}. {These observations also motivated reduced modelling approaches, with \citet{Doyle2007} explicitly distinguishing between static and flowing regions in a shallow-flow model of granular-column collapse. } Despite these observations, {many classical} scaling laws have been formulated in terms of the total mass and the global aspect ratio of the initial column, without explicitly accounting for how much of that mass is actually mobilised and how the static core constrains the geometry of the deposit. { Recent work has also incorporated quasi-static regions into runout scalings in inclined-to-horizontal configurations \citep{Chen2026}. %
In addition} to the mass partition, estimating how much of the initial gravitational potential energy is lost between the initial column and the final
deposit is crucial for developing reduced models with predictive capabilities. \cite{BalmforthKerswell2005} and \cite{Staron2005} have analysed the energetics of planar collapses and suggested that the sideways flow can be described by an effective restitution or friction at the base. {This motivates examining whether the static-core geometry can also constrain the final potential-energy budget.} Moreover, although analogies between laboratory collapses and long-runout landslides have been drawn \citep{Lajeunesse2006Mars}, quantitative comparisons of runout and dissipation between experiments and field events remain scarce.

{Building on these earlier observations of static inner cores, we revisit the dry collapse of axisymmetric granular columns and investigate how the measured geometry of the static region can be used as a geometrical closure linking deposit geometry, mass partition, and the final potential-energy budget. The latter is inferred from the barycentre of a slice of granular material of elementary azimuthal extension. }Section~\ref{sec:setup} presents the experimental set-up and image analysis methods. Section~\ref{sec:result} describes the deposit regimes, the scalings and a geometric model based on the observed static zone of constant angle. Section~\ref{sec:mass_energy} describes the mass partition and the motion of the slice barycentre, which allows for the estimation of an effective friction coefficient for the collapse from the dissipated energy.
Finally, Section~\ref{sec:field_comparison} compares our laboratory findings with Earth and Martian landslides to test the broader relevance of the developed framework.

\section{Experimental setup \label{sec:setup}}

The experimental set-up is sketched in figure~\ref{fig:Dispositif_exp}. A quarter of a cylindrical column of glass beads is initially prepared in the corner of a square glass tank of base area $1.2 \times 1.2~\mathrm{m}^2$ and height $0.30~\mathrm{m}$. The quarter granular column has an initial radius $R_0$ and height $H_0$, and contains a mass $M_0$ of {nearly monodisperse glass beads with a measured mean diameter $d=(4.96 \pm 0.06)$~mm} and density $\rho = (2.46 \pm 0.03)~\mathrm{g\,cm^{-3}}$. {The same bead material is used throughout the study, so that interparticle friction is not varied independently.} The glass beads are poured between the corner of the tank and a rigid quarter-cylindrical gate of 2 cm thick, and the free surface is then gently levelled with a flat ruler {without any additional tapping or vibration. The resulting initial bulk packing fraction $\phi_0=4\,M_0/(\rho\pi {R_0}^2H_0)$, depends slightly on the gate radius $R_0$ as expected from wall effects \citep{Suzuki2008} with smaller $\phi_0$ values at smaller $R_0$/d. For the largest gate radius $R_0 = 20~\mathrm{cm}$ corresponding to $R_0/d= 40$, the initial solid fraction is $\phi_0 \simeq 0.65$ corresponding to a dense packing. For the three other gate radii $R_0 = 10$, 5, and $3~\mathrm{cm}$, the initial solid fraction is smaller with the smallest value $\phi_0 \simeq 0.59$ for the smallest gate radius $R_0 = 3~\mathrm{cm}$ ($R_0/d = 6$) corresponding to the strongest confinement. Our initial granular columns thus range from relatively loose to dense packings depending on the system-to-grain-size ratio $R_0/d$. 

\begin{figure}
     \centering
        \includegraphics[width=0.6\textwidth]{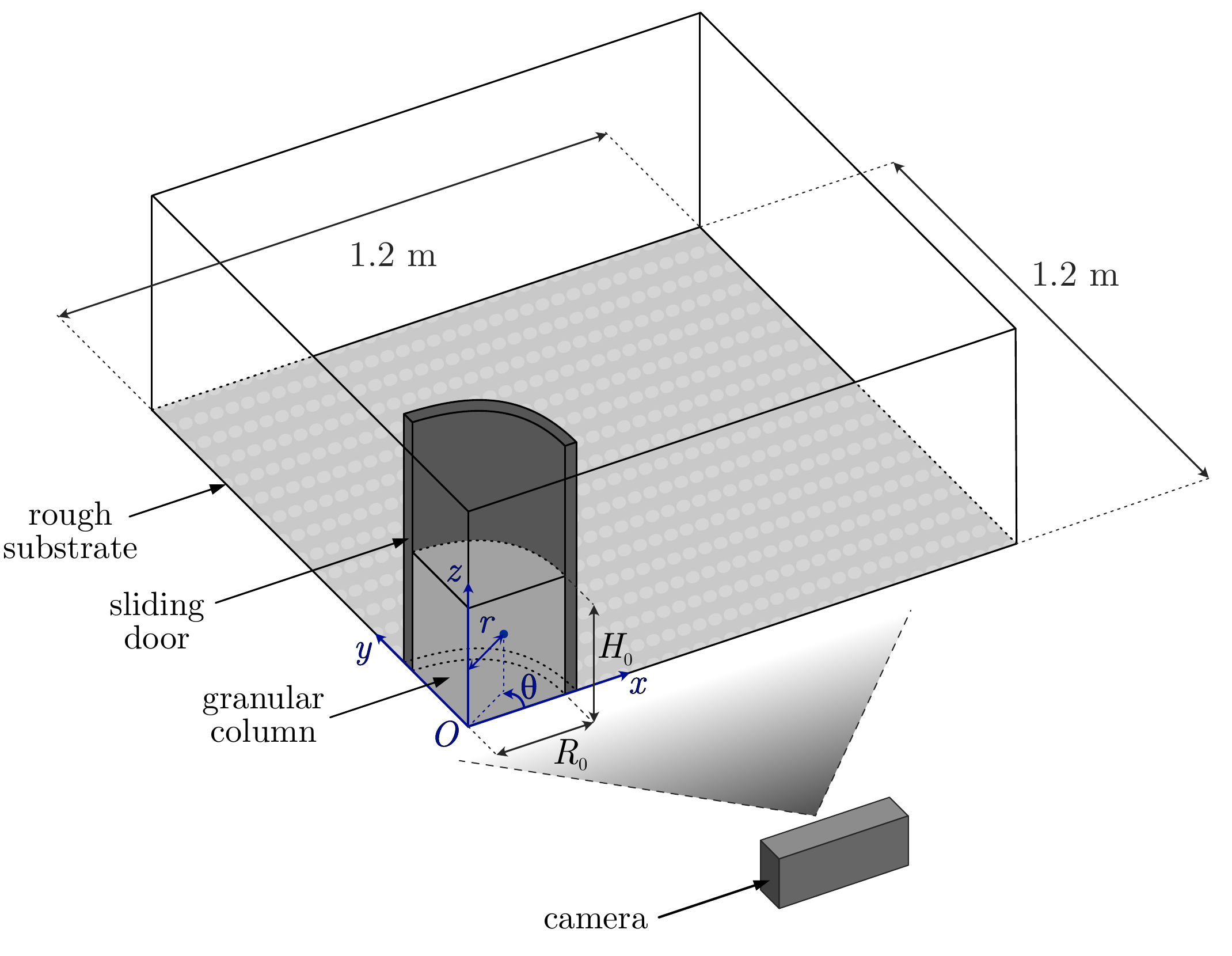}

         \caption{Schematic view of the experimental set-up.}
         \label{fig:Dispositif_exp}
\end{figure}

The gate is vertically guided by a rail and actuated by a motor-driven translation stage, while a lateral support frame prevents leakage and misalignment. The horizontal bottom of the glass tank is made rough by disposing some LEGO\textsuperscript{\textregistered} plates composed of cylindrical studs of diameter comparable with the glass bead diameter and height comparable with the bead half diameter. Indeed, the studs which form a regular square lattice of pitch $8~\mathrm{mm}$ are of diameter 4.8 mm and of height 1.7 mm.  To break the square periodicity, some additional small elements have been added over some initial studs without any periodicity, that increase both the stud diameter up to 7 mm and the stud height up to 3.2 mm. Such an irregular rough bottom suppresses systematic basal slip and unrestricted rolling of the bulk granular layer.} %
{When the gate is opened very slowly, at a typical velocity of 0.2 mm/s  much smaller than $(gd)^{1/2} \simeq 0.2$ m/s, where $g \simeq 9.8$ m/s$^2$ is the gravitational acceleration, the grain motion is very slow, so that the slope of the conical part formed by the final deposit should be close to the so-called angle of repose $\theta_r$ \citep{Sarlin2021PRE}. The measurements give $\theta_r = (23.9 \pm 0.7)^\circ$. This value is consistent with the value $\theta_r = (23.5 \pm 1.2)^\circ$ found by \cite{Sarlin2021PRE} for the same 5 mm glass beads but in a planar configuration, and consistent with the values reported in the literature for glass beads with a rotating drum setup \citep{FischerPRE2008}. 
We have also measured the sliding friction coefficient $\mu_w$ of our 5 mm glass beads on a glass wall. The measured value $\mu_w = 0.19 \pm 0.01$ is significantly smaller than $\tan\theta_r \simeq 0.45$, so that the friction effect at the wall is expected to be small in the present collapse experiments.}

{For the column collapse experiments, the gate is lifted at the maximal opening velocity of order $1~\mathrm{m\,s^{-1}}$ (average opening velocity $0.6$--$0.8~\mathrm{m\,s^{-1}}$, from the encoder), much larger than $(gd)^{1/2} \simeq 0.2$ m/s.}
In this range, we experimentally verified that the gate motion has no measurable influence on the collapse dynamics or the final deposit geometry, consistent with the criterion proposed in \cite{Sarlin2021PRE} for the collapse of 2D rectangular columns. { We checked also that the final deposits resulting from the column collapse remain axisymmetric, without any significant effect of the lateral confinement by the two lateral walls with solid friction \citep{Lajeunesse2005Slumping,Chen2026}. We therefore use the grain motion observed along one glass side wall, at the azimuthal position $\theta=0$, as representative of any meridional section $\theta$ in the cylindrical coordinates $(r,\theta,z)$, where $r$ is the radial coordinate measured from the corner of the tank, $\theta$ is the azimuthal angle measured from the front side wall, and $z$ is the vertical coordinate, as shown in figure~\ref{fig:Dispositif_exp}.}

The motion of the column and the subsequent granular flow are recorded with a digital camera (Nikon D3300) operating at a frame rate of {50 fps} and a resolution of $1920 \times 1080$ pixels. The camera is mounted on a tripod at a distance of more than $2~\mathrm{m}$ from the tank, with its optical axis normal to the front side wall and centred on the initial position of the gate to minimise parallax. {The relatively large camera distance also reduces perspective variations across the field of view. The spatial resolution varies from approximately $0.4$ to $1.4~\mathrm{mm\,pixel^{-1}}$, with a mean value of about $0.81~\mathrm{mm\,pixel^{-1}}$. A $5$-mm grain therefore spans approximately $3.6$--$12.5$ pixels, with about $6$ pixels per grain on average.} The set-up is illuminated from above and from the front with {three white-light LED lamps} to maximise the contrast between the grains and the background. Spurious reflections are reduced by covering the surroundings with black fabric and regularly cleaning the glass walls.

\begin{figure}
     \begin{subfigure}[b]{0.6\textwidth}
         \centering
         \includegraphics[width=\textwidth]{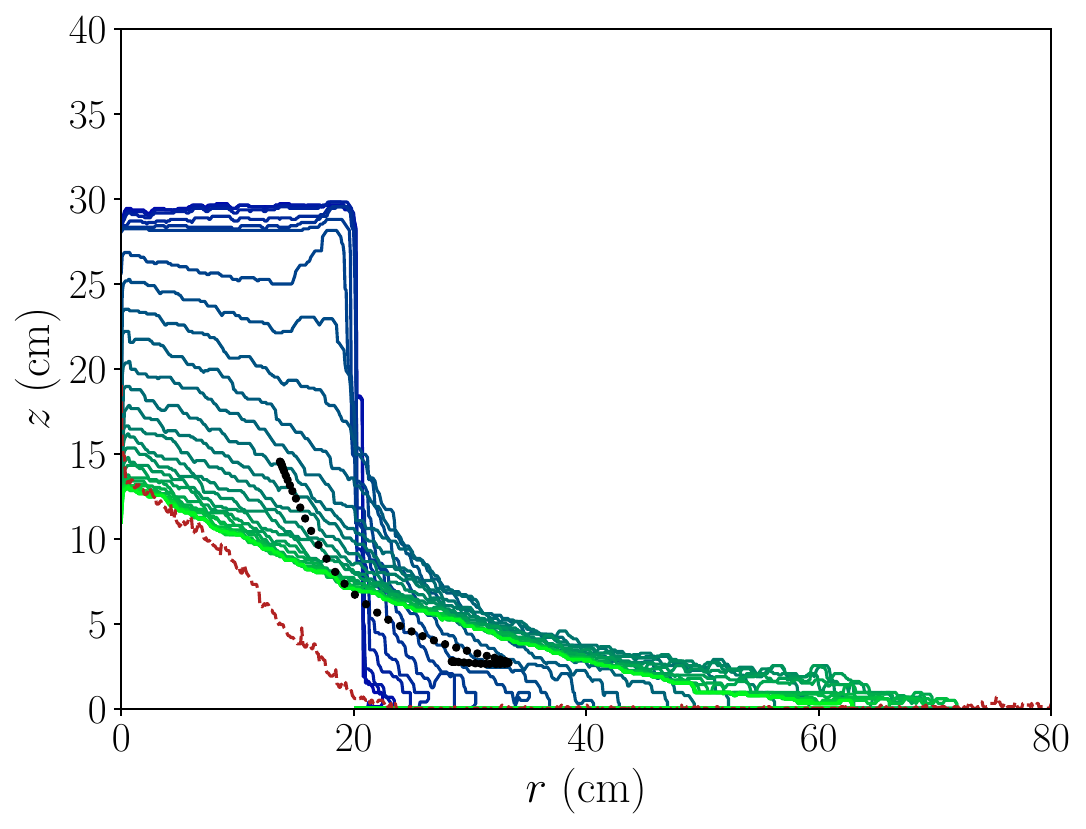}
         \caption{}
         \label{fig:tracking}
     \end{subfigure}
     \hfill
     \begin{subfigure}[b]{0.45\textwidth}
         \centering
         \includegraphics[width=\textwidth]{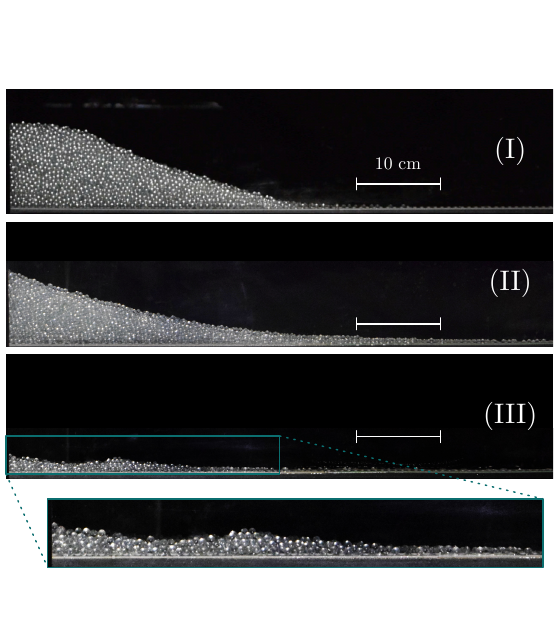}
         \caption{}
         \label{fig:Depots}
     \end{subfigure}
     \hfill
    \caption{(a) Successive interfaces $h(r,t)$ in solid lines from blue to green for the collapse of an initial column of height $H_0 = 30$ cm and radius $R_0 = 20$ cm ($a = 1.5$). To make it easier to read, only one in two images is shown. The black circles correspond to the successive positions of the barycentre $G$ of all the grains contained in an elementary slice of vanishing extension ${\rm d}\theta$, defined in \S\ref{sec:Barycentre}. {The red dashed line corresponds to the static--flowing interface $h_s(r)$, defined in \S\ref{sec:angles}.} (b) Photographs of the final deposits for three representative collapses, illustrating the three regimes observed as the aspect ratio $a=H_0/R_0$ increases: regime~I, $H_0=7~\mathrm{cm}$ and $R_0=20~\mathrm{cm}$ ($a=0.35$); regime~II, $H_0=30~\mathrm{cm}$ and $R_0=10~\mathrm{cm}$ ($a=3$); and regime~III, $H_0=50~\mathrm{cm}$ and $R_0=3~\mathrm{cm}$ ($a=17$).}
    \label{fig:effondrements}
\end{figure}

After acquisition, each movie is processed with a custom MATLAB routine. The sequence is first converted into individual grayscale images. { The time origin, $t=0$~s, is defined as the instant when the gate leaves the substrate.} We denote by $t_\infty$ the time beyond which the final deposit no longer evolves, typically of the order of a second. To extract the interface, each frame is partitioned into small subregions and thresholded to retain only the bright pixels corresponding to grains. The resulting binary image is then smoothed by morphological erosion and dilation to remove isolated pixels and ejecta. The interface height $h(r,t)$ between grains and air from the horizontal bottom at $z = 0$ is obtained as the boundary of the connected region of ``grain'' pixels for each radial position $r$ from the corner of the tank at $r = 0$. An example of the successive interfaces $h(r,t)$ tracked during the collapse of an initial granular column of aspect ratio $a = 1.5$ is shown in figure~\ref{fig:effondrements}(a). One can see that a thin zone of grains close to the gate is
{transiently entrained upwards} by the lifting of the gate at the beginning of the collapse, but {these grains rapidly fall back and catch up with the rest of the interface.} The top of the column essentially falls vertically, whereas the bottom of the column spreads radially with an increasing radius $R(t)$ at the bottom ($z = 0$). In this example, the final deposit is conical.

The control parameters of the experiments are the initial height $H_0$ and radius $R_0$, resulting in the initial aspect ratio $a = H_0/R_0$ of the granular column, which is varied over nearly two decades by adjusting the mass of grains $M_0$ and the gate radius $R_0$. The ranges of parameters explored are summarised in table~\ref{tab:params}. Uncertainties regarding the control parameters are on the order of the grain size, namely 0.5 cm. %
{
After the collapse, the final height of the deposit $H_\infty=h(0,t_\infty)$ is measured along the $z$-axis of symmetry at $r=0$. The final radius $R_\infty=R(t_\infty)$ is measured from the symmetry axis
$r=0$ to the outer edge of the deposit at the rough substrate ($z=0$), defined as the first radial position where the grain layer becomes discontinuous over
the roughness elements, with $h(R_\infty,t_\infty)=0$. In the following, we will mainly focus on this final spreading distance and not on the runout distance $\Delta R_\infty = R_\infty-R_0$ beyond the initial footprint $R_0$, which is another commonly used measurement for granular collapse.} The uncertainty in $R_\infty$ is typically a few centimetres depending on the size of the deposit, but the relative uncertainty is always less than or equal to 10\%. Several experiments with the same parameters are performed to assess reproducibility, and the dispersion of the measured quantities is found to be smaller than the error bars.

\begin{table}
  \centering
  \begin{tabular}{lcr}
    \hline
    Parameter & Notation & Range \\
    \hline
    Mass of grains & $M_0$ & $0.05$--$19~\mathrm{kg}$ \\
    Initial height & $H_0$ & $5$--$60~\mathrm{cm}$ \\
    Initial radius & $R_0$ & $3$, $5$, $10$, $20~\mathrm{cm}$ \\
    Aspect ratio & $a = H_0/R_0$ & $0.25$--$21.5$ \\
    \hline
  \end{tabular}
  \caption{Summary of the experimental control parameters and the corresponding ranges explored.}
  \label{tab:params}
\end{table}

\section{Anatomy of a fall \label{sec:result}}

\subsection{Scaling laws for the deposit \label{sec:res_scaling}}

\begin{figure}
    \centering
    \begin{subfigure}{0.48\textwidth}
        \includegraphics[width=\textwidth]{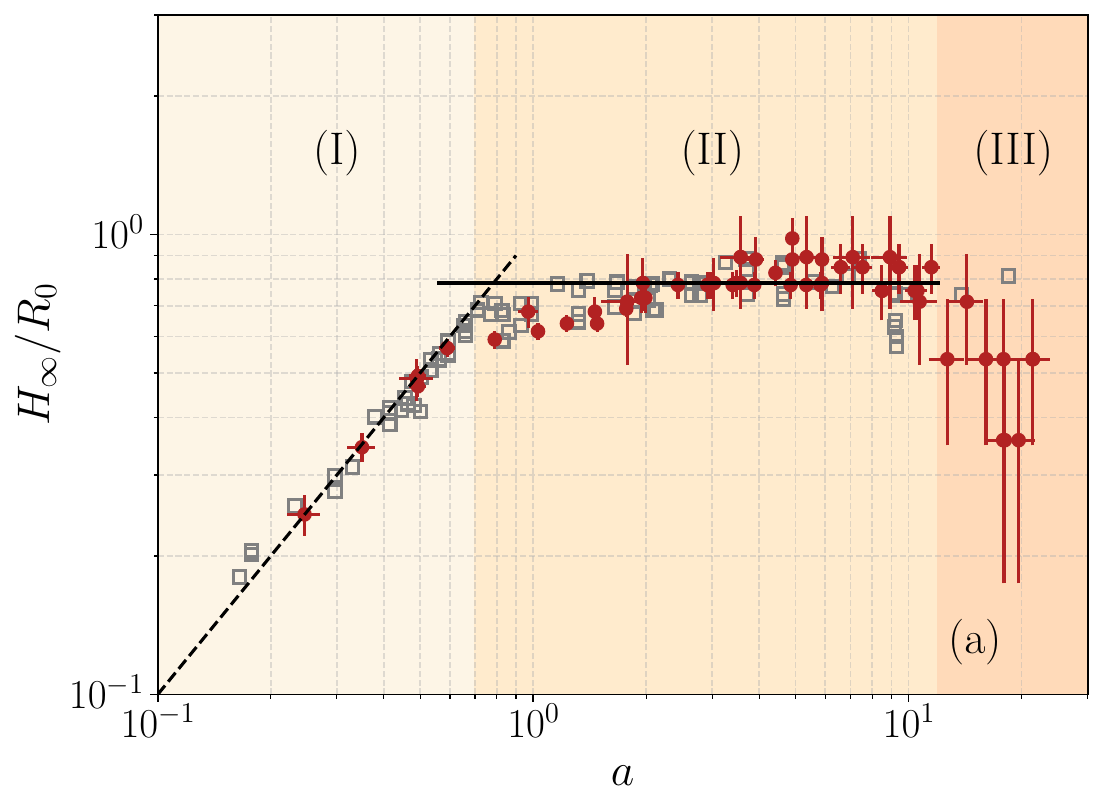}
        \label{fig:Hf/R0}
    \end{subfigure}
    \hfill
    \begin{subfigure}{0.48\textwidth}
        \includegraphics[width=\textwidth]{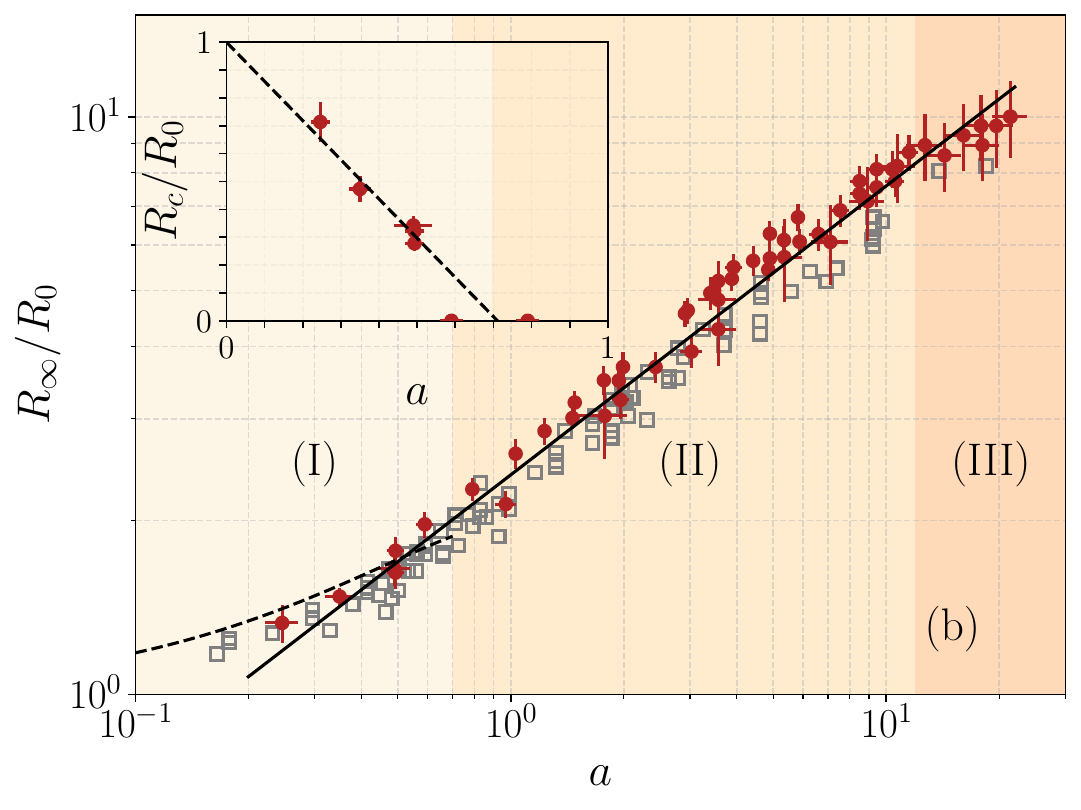}
        \label{fig:Rf/R0}
    \end{subfigure}
    \\
    \begin{subfigure}{0.48\textwidth}
        \includegraphics[width=\textwidth]{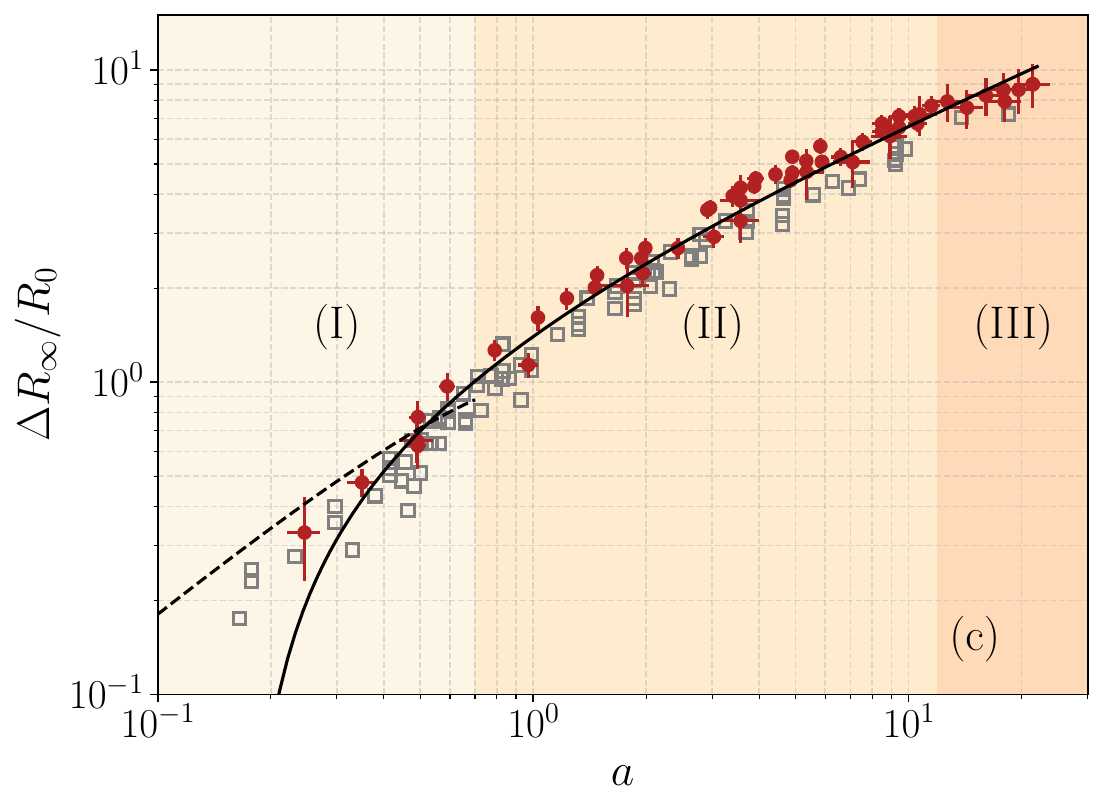}
        \label{fig:dRf_R0}
    \end{subfigure}
    \hfill
    \begin{subfigure}{0.48\textwidth}
        \includegraphics[width=\textwidth]{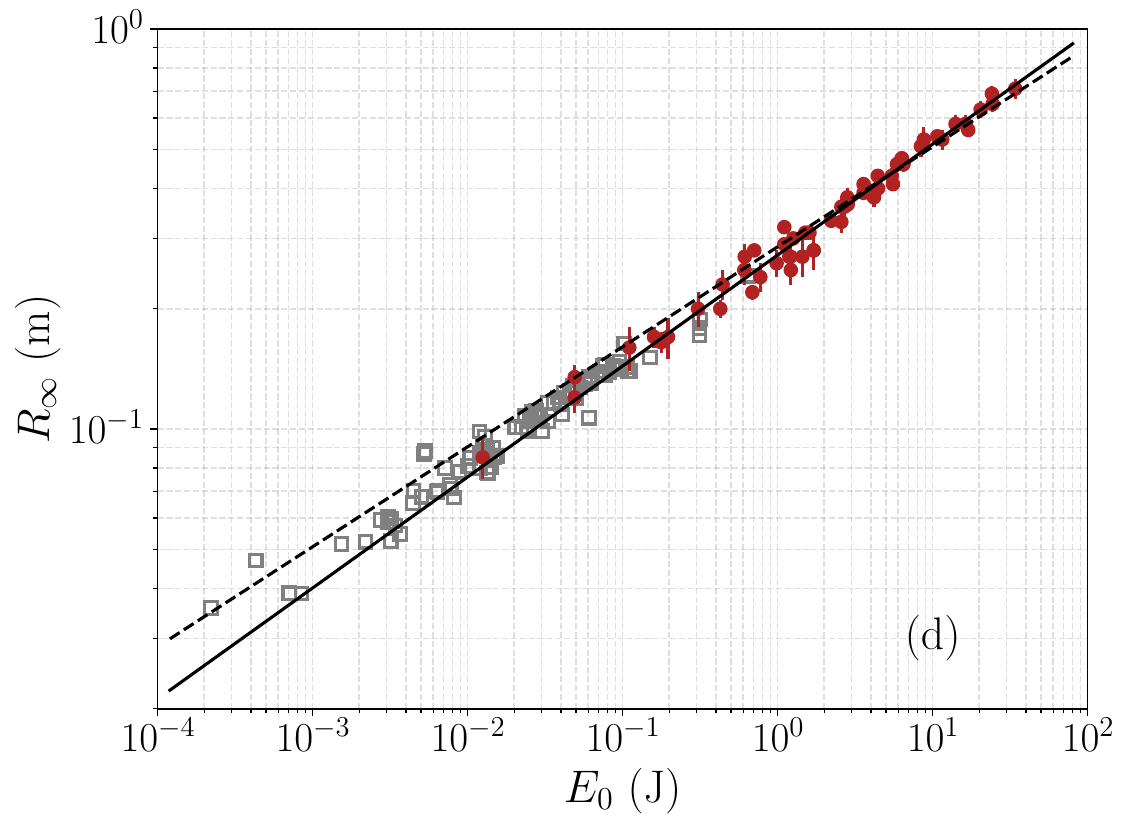}
        \label{fig:Rf_vs_E0}
    \end{subfigure}
    \caption{(a) Normalised final height $H_\infty/R_0$ as a function of the aspect ratio $a=H_0/R_0$. The dashed line corresponds to $H_\infty/R_0=a$ (regime~I), and the horizontal solid line to the plateau value $H_\infty/R_0=0.79$ in regime~II. (b) Normalised final radius $R_\infty/R_0$ as a function of $a$. The solid line corresponds to the regime-II fit $R_\infty/R_0=D a^{1/2}$, with $D=2.4$. The dashed line corresponds to the regime-I truncated-cone expression \eqref{eq:Rf_regimeI}, fitted to the regime-I runout data with the value 0.52. Inset: normalised truncation radius $R_c/R_0$ as a function of the aspect ratio $a$, where the dashed line corresponds to the best fit of the data by equation (\ref{eq:Rc}) with $\tan \theta_2 = 0.71$, corresponding to a geometric estimate of $a_c$ from the truncation-radius data. (c) Normalised runout $\Delta R_\infty/R_0$ as a function of $a$. The solid line corresponds to $\Delta R_\infty/R_0=2.4 \,a^{1/2} -1$. {This curve follows directly from the fit of the present $R_\infty/R_0$ data shown in panel (b), through $\Delta R_\infty/R_0=R_\infty/R_0-1$.} The dashed line corresponds to the regime-I expression \eqref{eq:Rf_regimeI} minus one, fitted to the regime-I runout data with the value 0.52. (d) Final radius $R_\infty$ as a function of the initial potential energy $E_0=M_0 g H_0/2$ of a quarter-cylinder column. { The solid line corresponds to the empirical fit $R_\infty=\beta E_0^{0.28}$ given by equation~\eqref{eq:Rf_E0_emp}, with $\beta=0.27\,\mathrm{m\,J^{-0.28}}$, {whereas the dashed line corresponds to a fit with the exponent fixed to $1/4$, $R_\infty=\alpha E_0^{1/4}$, yielding $\alpha=0.286\pm0.001\,\mathrm{m\,J^{-1/4}}$.} The symbols correspond to experimental data (\raisebox{-0.2ex}{\Large\textcolor{BrickRed}{$\bullet$}}) of the present study and (\textcolor{Gray}{$\square$}) of \cite{Lajeunesse2004Spreading}.}}
    \label{fig:Quatuor_depot}
\end{figure}

We first characterise the geometry of the final deposit in terms of its height $H_\infty$ and horizontal extent $R_\infty$. Following previous studies on the collapse of granular columns \citep{Lajeunesse2004Spreading,Lube2004}, we focus on the normalised final height $H_\infty/R_0$ and radius $R_\infty/R_0$ as functions of the aspect ratio $a = H_0/R_0$. Our measurements for $d= 5$ mm glass beads (\raisebox{-0.2ex}{\Large\textcolor{BrickRed}{$\bullet$}}) are reported in figures \ref{fig:Quatuor_depot}(a,b) together with the data obtained by \cite{Lajeunesse2004Spreading} (\textcolor{Gray}{$\square$}) for a full column of $d = 1$ mm glass beads. Within experimental scatter, the two data sets collapse very well, {showing that the global deposit dimensions measured in the present quarter-cylinder geometry are consistent with those reported for full cylindrical columns. The small differences may result from finite-size effects associated with the different values of $R_0/d$, as well as from differences in basal roughness.}

\bigbreak

For small aspect ratios, the final deposit has the shape of a truncated cone [regime~I in figure~\ref{fig:effondrements}(b)], where the top remains at the initial height: $H_\infty = H_0$. In this regime, the normalised height therefore increases linearly with the aspect ratio, $H_\infty/R_0 = a$, and all data collapse on the dashed line shown in figure~\ref{fig:Quatuor_depot}(a). As the aspect ratio increases, the truncated part of the deposit characterised by the radius $R_c$ decreases progressively toward zero at $a_c = 0.7 \pm 0.1 $ as shown in the inset of figure~\ref{fig:Quatuor_depot}(b). The truncation radius $R_c$ is defined as the first radial position (from the axis) where the final free-surface height drops below $0.95\,H_0$, \textit{i.e.} $h(r=R_c,t_\infty)=0.95\,H_0$. At this specific aspect ratio, $a_c = 0.7 \pm 0.1 $, the deposit looks like a cone, with a final height $H_\infty$ still equal to $H_0$. 

When the aspect ratio exceeds the critical value $a_c$, the deposit becomes a smoothly incurved cone [regime~II in figure~\ref{fig:effondrements}(b)]. In this regime, the increase of the normalised height, $H_\infty/R_0$, is strongly reduced and tends towards a nearly constant value.  As in previous axisymmetric experiments, the behaviour of regime II can either be described by a power law $H_\infty \propto a^{1/6}$ with a very small exponent as proposed by \cite{Lube2004}, or by a plateau value as suggested by \cite{Lajeunesse2004Spreading}. A power-law fit to our data in the range $0.8 \lesssim a \lesssim 6$ gives $H_\infty/R_0 \simeq 0.65 a^{0.16}$, in good agreement with \cite{Lube2004}. However, over the broader range $0.8 \lesssim a \lesssim 12$, the data are well described by a constant value, with a mean normalised height $ H_\infty/R_0  = 0.79 \pm 0.08$, shown as the horizontal solid line in figure~\ref{fig:Quatuor_depot}(a). In the following, we adopt this simpler ``plateau'' description for regime~II. For even larger aspect ratios, typically $a \gtrsim 12$, the deposit no longer resembles a cone but consists of two shallow lobes [regime~III in figure~\ref{fig:effondrements}(b)]. In this regime, the initial granular column is very slender, and the collapse resembles a granular jet impacting a rough surface and generating a transient granular jump \citep{boudet2007}. This leads to a significant decrease of $H_\infty/R_0$ with $a$ in regime III, as can be seen in figure \ref{fig:Quatuor_depot}(a), and to another bump at non-zero $r$, as can be observed in the bottom image of figure~\ref{fig:effondrements}(b). {In this impact-dominated regime, the maximum height of the final
deposit is reached away from the symmetry axis, so that $H_\infty=h(0)$ no longer provides a representative measure of the deposit height. Consequently, the conical static-core construction used in regimes~I and II is no longer
geometrically appropriate. } %
{Because regime-III deposits are only a few grain diameters thick, we did not attempt to characterise this regime quantitatively in the present study. A similar transition to regime III at high $a$ was reported by \citet{Staron2005} but not by
\citet{Lajeunesse2004Spreading}. In the present experiments, regime~III occurs only for the smallest system-to-grain-size ratio, $R_0/d=6$, suggesting that finite-size effects may be important \citep{Warnett2014,Cabrera2021,Man2021FiniteSize}.}

Based on the above observations, we summarise the evolution of the final height of the deposit in regimes I and II by the following piecewise scaling laws:
\begin{subequations}
\label{eq:Hf_piecewise}
\begin{align}
\frac{H_\infty^{\mathrm{I}}}{R_0} &= a 
 \quad  \text{for } a \leqslant a_c,
\label{eq:Hf_regimeI}\\
\frac{H_\infty^{\mathrm{II}}}{R_0} &= C
\quad  \text{for } a_c \leqslant a \lesssim 12,
\label{eq:Hf_regimeII}
\end{align}
\end{subequations}
with $C=0.79\pm 0.08$. This plateau value is consistent with previous studies, which reported plateau values of $H_\infty/R_0 \simeq 0.74$ and $1$ in \cite{Lajeunesse2004Spreading} and \cite{Lube2004}, respectively. %
{The piecewise description would be exactly continuous if $a_c=C$. Here, however, these two quantities are determined independently: $a_c=0.7\pm0.1$ from the disappearance of the truncation radius $R_c$
[inset of figure~\ref{fig:Quatuor_depot}(b)] in regime I, and $C=0.79\pm0.08$ from the mean height plateau in regime~II [figure~\ref{fig:Quatuor_depot}(a)]. Their values are compatible within experimental uncertainty, so the small
difference between them should not be interpreted as a physical discontinuity. In the following, we retain $a_c=0.7\pm0.1$ as the transition value, and will consider $C \simeq a_c$ in the modelling.}

In contrast with the normalised final height, the normalised final radius $R_\infty/R_0$ does not exhibit a clear change of behaviour {across the different regimes}
[figure~\ref{fig:Quatuor_depot}(b)]. { In particular, the transition from regime~I to regime~II is progressive: as $a$ increases in regime~I, the truncation radius $R_c$ decreases continuously and vanishes at $a=a_c$, so that the truncated deposit characteristic of regime I turns to the conical deposit characteristic of regime II, and $R_\infty/R_0$ increases continuously with $a$ in regimes I and II, and also in regime III, without any discontinuity.
}
{Most of the data} are quite well captured by the scaling law 
\begin{equation}
\frac{R_\infty}{R_0}=D\,a^{1/2},
\label{eq:Rf_scaling}
\end{equation}
with $D=2.40 \pm 0.03$ (uncertainty estimated by a Monte Carlo procedure). Such a power scaling with exponent 1/2 will be derived in regime II from a simple geometric
model developed in \S\ref{sec:geom_model}, along with a theoretical prediction for the numerical coefficient $D$. For the smallest aspect ratios of regime~I, the measured normalised lateral radius $R_\infty/R_0$ exhibits some significant deviation from this simple power scaling, and a modelling based on a truncated cone geometry also developed in \S\ref{sec:geom_model} predicts such a
progressive deviation.

Many previous works have focused on the run-out distance $\Delta R_\infty = R_\infty - R_0$ and { have described the evolution of $\Delta R_\infty/R_0$ with $a$ with two empirical power-law regimes} \citep{Lajeunesse2004Spreading,Lube2004}. For comparison, we also plot $\Delta R_\infty/R_0$ in figure~\ref{fig:Quatuor_depot}(c). Since $\Delta R_\infty/R_0 = R_\infty/R_0 - 1$, the square-root scaling~\eqref{eq:Rf_scaling} implies
\begin{equation}
\frac{\Delta R_\infty}{R_0} = D\,a^{1/2} - 1,
\end{equation}

\noindent which describes quite well all the data as illustrated by the solid line in figure~\ref{fig:Quatuor_depot}(c) except for the smallest $a$ values of regime I, where there is some significant deviation that will be modelled in \S\ref{sec:geom_model} as shown by the dashed line. %
Because $\Delta R_\infty/R_0$ contains no additional information beyond $R_\infty/R_0$, and since $R_\infty$ enters naturally in the geometric construction developed below, we will focus on $R_\infty/R_0$ in the following.

Beyond its dependence on the aspect ratio, one can relate the spreading distance $R_\infty$ to the initial potential energy $E_0=gM_0H_0/2$ of the quarter column. Figure~\ref{fig:Quatuor_depot}(d) shows the evolution of $R_\infty$ with $E_0$ for our experiments, together with the data of \cite{Lajeunesse2004Spreading}. Because their experiments involve a full cylindrical column, while ours correspond to a quarter cylinder, their initial energies have been divided by $4$ for consistency. All data in figure~\ref{fig:Quatuor_depot}(d) collapse onto a master curve, which is well captured by the power-law scaling
\begin{equation}
R_\infty = {\beta} E_0^{\,0.28},
\label{eq:Rf_E0_emp}
\end{equation}

\noindent { where $\beta=0.27\, \mathrm{m\,J^{-0.28}}$ is the dimensional fitted
prefactor.} The solid-line fit~\eqref{eq:Rf_E0_emp} with the free-fitted exponent 0.28, which is close to 1/4, describes the experimental points to better than $10\%$ on average, showing that $E_0$ provides a compact parametrisation of the runout over the explored range.
A power scaling of $R_\infty$ with $E_0$ with a power exponent 1/4 can be deduced directly from the power scaling of $R_\infty$ with $a$ with the power exponent 1/2 given by equation~\eqref{eq:Rf_scaling}. Indeed, for a quarter-cylinder of initial radius $R_0$ and height $H_0$, the initial mass is $M_0=(\pi/4)\,\phi_0\,\rho\,R_0^2H_0$, so that the initial potential energy $E_0=M_0gH_0/2$ reads
\begin{equation}
E_0 = \frac{\pi \phi_0 \rho g}{8}\,R_0^2 H_0^2.
\label{eq:E0}
\end{equation}
The combination of equations~\eqref{eq:Rf_scaling} and~\eqref{eq:E0} then yields the following energy-based scaling for the final radius:
\begin{equation}
R_\infty = \alpha\,E_0^{1/4},
\label{eq:Rf_E0_th}
\end{equation}
with a predicted power exponent $1/4$ and a predicted coefficient $\alpha=D(\pi\phi_0\rho g/8)^{-1/4}$. Taking the measured values of $D$ and $\rho$ with $\phi_0 = 0.62 \pm 0.03$ gives $\alpha=(0.274\pm0.007)\,\mathrm{m\,J^{-1/4}}$, close to the value $\alpha=(0.286\pm0.001)\,\mathrm{m\,J^{-1/4}}$ obtained from the best fit to the data using equation~\eqref{eq:Rf_E0_th}. The resulting fit, which is shown as the dashed line in figure~\ref{fig:Quatuor_depot}(d) close to the solid line from the free empirical fit ~\eqref{eq:Rf_E0_emp}, captures our measurements to better than $10\,\%$ on average and reproduces equally well the data of \cite{Lajeunesse2004Spreading}. This initial energy parametrisation is not expected to be sufficient in regime~I, since $E_0$ does not fix the aspect ratio and one rather expects $R_\infty^{\mathrm{I}}(E_0,a)$. But in practice, the $E_0^{1/4}$ law provides a remarkably good description of the runout data, even at low aspect ratios. The small difference between the exponents $0.28$ and $1/4$ lies within the uncertainty associated with the experimental scatter and the limited range of accessible energies. This form may be particularly convenient for field applications, where $E_0$ (via volume and fall height) is often easier to constrain than idealised geometric parameters $H_0$ and $R_0$.

\subsection{Internal structure of the collapse}
\label{sec:angles}

\begin{figure}
    \centering
    \includegraphics[width=0.9\textwidth]{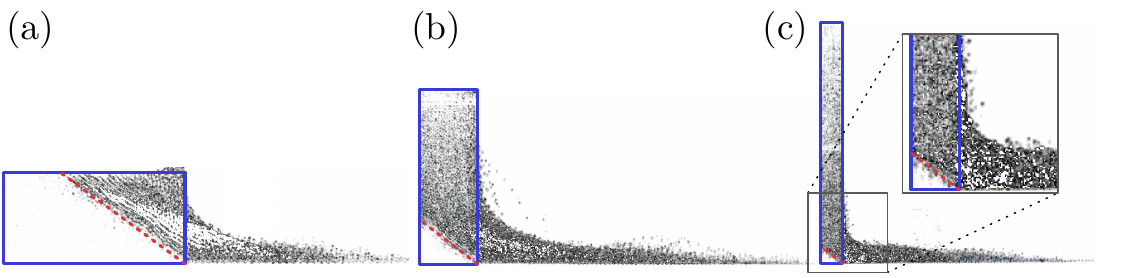}
    \caption{Maps of mobile and immobile regions obtained from frame-to-frame image subtraction {over the entire collapse duration} for three representative collapses. Pixels corresponding to grains that have moved at least once during the collapse are shown in black, whereas grains that remain immobile throughout are shown in white. The blue square indicates the outline of the initial column. (a) Regime~I ($H_0=10~\mathrm{cm}$, $R_0=20~\mathrm{cm}$, $a=0.5$): the red line (\textcolor{Red}{-{}-{}-}) marks the interface of the static region, joining the first point (from the axis) along the final free-surface profile where $h(r,t_\infty)<H_0$ to $(r=R_0,\,z=0)$. (b,c) Regime~II for $H_0=30~\mathrm{cm}$, $R_0=10~\mathrm{cm}$ ($a=3$) and $H_0=60~\mathrm{cm}$, $R_0=5~\mathrm{cm}$ ($a\simeq12$), respectively: the red line (\textcolor{Red}{-{}-{}-}) highlights the boundary of the static inner cone, joining $(r=0,\,z=H_\infty)$ to $(r=R_0,\,z=0)$.}
    \label{fig:Mg}
\end{figure}

Beyond the global scalings discussed in \S\ref{sec:res_scaling}, well-defined static and mobile regions delimited by characteristic slope angles are observed. We now describe these geometric features, which will underpin the simple geometric model introduced in \S\ref{sec:geom_model}, as well as the mass partition and final energy budget discussed in \S\ref{sec:mass_energy}.

To visualise which parts of the column actually move during the collapse, we subtract each frame of a movie from the previous one, so that pixels that have changed intensity between two frames correspond to moving grains. Summing all these difference images over the whole duration of the collapse and thresholding the result yields a binary map that separates mobile from static zones. {The threshold was chosen above the frame-to-frame intensity fluctuations measured in a region known to remain static, and we checked that a small change of the threshold does not modify the results significantly. This cumulative procedure identifies whether a region moves at least once during the collapse.} For clarity, figure~\ref{fig:Mg} shows the negative of this binary map: black pixels mark regions where intensity changes were detected at least once during the collapse (indicating grain motion), whereas white pixels correspond to regions that remain unchanged throughout (immobile zone). The outline of the initial column is indicated by a blue rectangle in figure~\ref{fig:Mg}.

In regime~I, when the aspect ratio is small enough ($a < a_c$), as for the case corresponding to $a = 0.5$ shown in figure~\ref{fig:Mg}(a), only a small part of the initial column far from the axis of symmetry is flowing, whereas a large part attached to the axis of symmetry remains essentially immobile. This static part appears as the white region below the red dashed straight line in figure~\ref{fig:Mg}(a) and corresponds to an inner truncated cone of radius $R_0$ and height $H_0$. In regime~II, when the aspect ratio is larger ($a > a_c$), a large fraction of the column is mobilised, while {an inner conical core of grains} remains essentially immobile at the base near the symmetry axis. This can be seen for the case corresponding to $a\simeq 3$ in figure~\ref{fig:Mg}(b). The static part appearing below the red dashed straight line in figure~\ref{fig:Mg}(b) corresponds to a cone of radius $R_0$ and height $H_\infty$. For an even larger aspect ratio, e.g., for the case of $a \simeq 12$ shown in figure~\ref{fig:Mg}(c) close to the transition to regime~III, most of the column is flowing, {while a small white region is still visible and corresponds to a static conical region of radius $R_0$ and height $H_\infty$. In regime~II, the fraction of the initial mass contained in the static region decreases markedly as $a$ increases, as quantified in \S\ref{section:mass-partition}.} Note that the slope angle of the static part appears to be very similar in the three pictures of figure~\ref{fig:Mg}. The observation of a static inner cone surrounded by a flowing outer granular region is fully consistent with the previous observations of \cite{Lajeunesse2004Spreading}, \cite{Lube2004}, and \cite{Man2021} for the collapse of cylindrical granular columns. In figure~\ref{fig:effondrements}(a), where the successive interfaces $h(r,t)$ are drawn in solid lines from blue to green for one typical collapse in regime II, the static interface $h_s(r)$, which is represented schematically by a red dashed line in figure \ref{fig:Mg}, is also drawn in red dashed line in figure~\ref{fig:effondrements}(a). One can see in this example that $h_s(r)$ is indeed very close to a straight line from $(z =H_\infty, r = 0)$ to $(z= 0,r = R_0)$.

\begin{figure}
    \centering
    \begin{subfigure}[b]{0.45\textwidth}
         \centering
         \includegraphics[width=\textwidth]{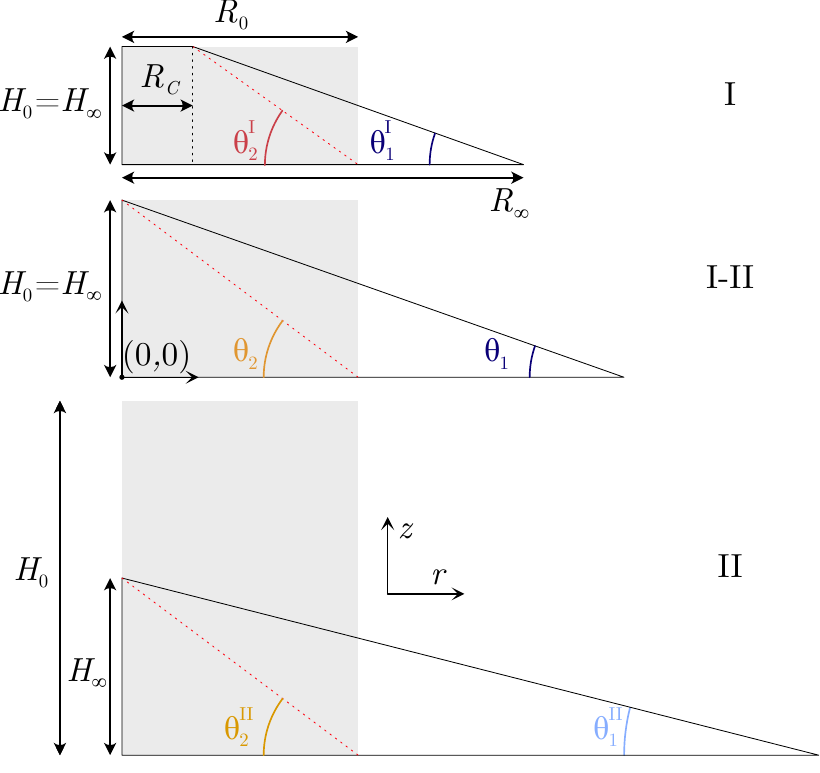}
         \caption{}
         \label{fig:Schéma_angles}
     \end{subfigure}
     \hfill
     \begin{subfigure}[b]{0.45\textwidth}
         \centering
         \includegraphics[width=\textwidth]{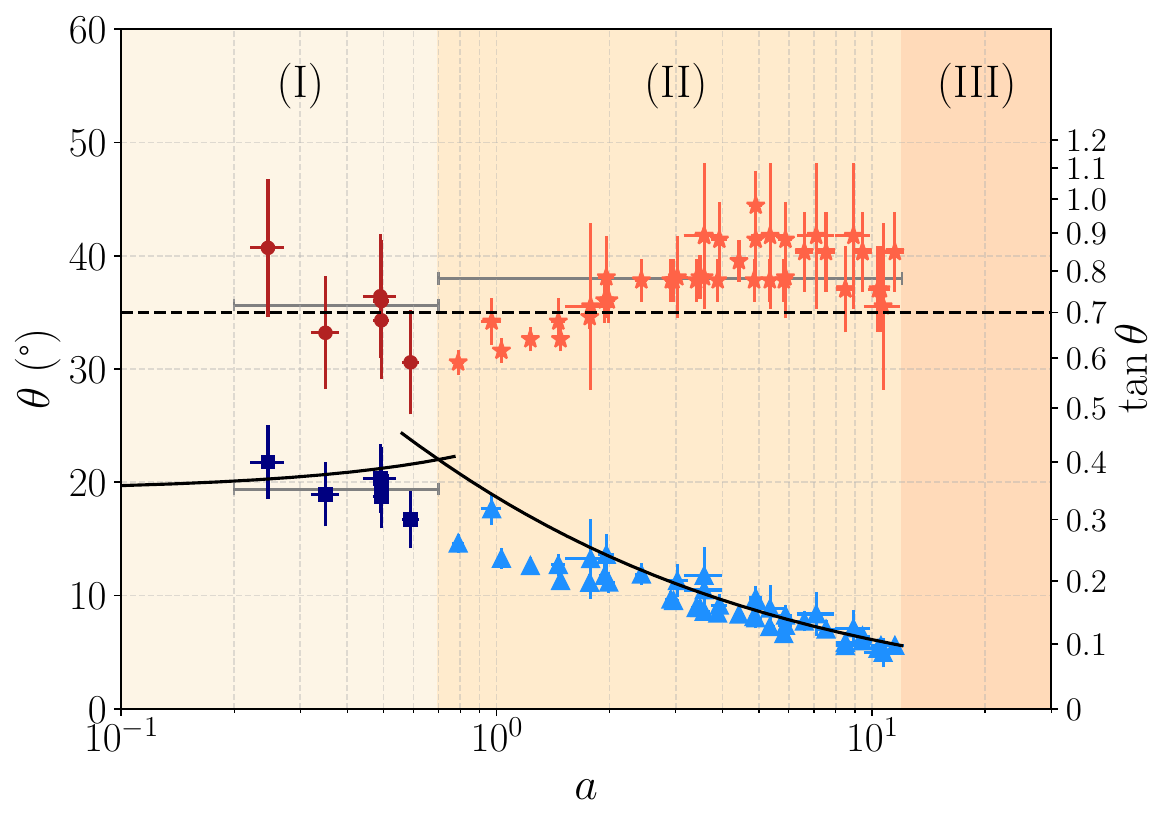}
         \caption{}
         \label{fig:Mesures_angles}
     \end{subfigure}
     \hfill
    \caption{(a) Schematic representation of the deposit geometry in regimes~I and II. The grey area indicates the initial shape of the column, and the angles $\theta_2$ and $\theta_1$ are defined as in the text. (b) Evolution of the angles $\theta_2$ and $\theta_1$ with the aspect ratio $a$ in regimes~I and II. The horizontal grey lines (\textcolor{Black!50}{---}) indicate the mean values of the measured angles. The dashed line corresponds to $\arctan(a_c)$. The solid lines correspond to equations \eqref{eq:theta_1_I} and \eqref{eq:theta_1_II} with $a_c = 0.7$. Symbols and colours are defined in table~\ref{tab:theta} and in panel~(a).}
    \label{fig:Modele}
\end{figure}

These qualitative observations motivate a compact geometric description of the deposits with two characteristic angles measured from the horizontal, as schematically illustrated in figure \ref{fig:Modele}(a). We denote by $\theta_1$ the mean angle of the outer free surface of grains that have flowed, and by $\theta_2$ the angle of the interface separating the static inner cone from the surrounding flowing region.  In regime~I, the final deposit is idealised as a truncated cone of angle $\theta_1^{\mathrm{I}}$ surrounding a static inner truncated cone of angle $\theta_2^{\mathrm{I}}$ [figure~\ref{fig:Modele}(a) (top)]. The static interface at angle $\theta_2^{\mathrm{I}}$ joins the upper corner of the static region at $(r = R_c,z = H_0)$ to the base of the initial column at $(r = R_0,z = 0)$, so that $\tan\theta_2^{\mathrm{I}} = H_0/(R_0-R_c)$, while the outer slope at angle $\theta_1^{\mathrm{I}}$ connects the same upper point $(r = R_c,z = H_0)$ to the front of the deposit at the bottom $(r = R_\infty,z = 0)$, so that $\tan\theta_1^{\mathrm{I}} = H_0/(R_\infty^{\mathrm{I}}-R_c)$. In regime~II, the deposit is represented as a full cone [figure~\ref{fig:Modele}(a) (bottom)], with a static inner cone of radius $R_0$ and height $H_\infty$ delimited by the slope of angle $\theta_2^{\mathrm{II}}$, so that $\tan\theta_2^{\mathrm{II}} = H_\infty^{\mathrm{II}}/R_0$, and a flowing outer shell extending out to the final radius $R_\infty$ with the free-surface angle $\theta_1^{\mathrm{II}}$, so that $\tan\theta_1^{\mathrm{II}} = H_\infty^{\mathrm{II}}/R_\infty^{\mathrm{II}}$.  

Using the measured shape of the deposits and the simplified geometries sketched in figure~\ref{fig:Modele}(a), we can infer the angles $\theta_1$ and $\theta_2$ {from the measured geometrical quantities, using the expressions summarised in table~\ref{tab:theta},} for both regimes I and II. The resulting  angles $\theta_1$ and $\theta_2$ are plotted in figure~\ref{fig:Modele}(b). To leading order, the inner angle $\theta_2$ is nearly independent of $a$ in both regimes~I and II, with a mean value $\theta_2 = (37 \pm 4)^\circ$. { More precisely, the values of $\theta_2$ in the two regimes are $\theta_2^{\rm I}=(35.6 \pm 3.0)^\circ$ and $\theta_2^{\rm II}=(38.2 \pm 2.8)^\circ$ as reported in table~\ref{tab:theta}.} This confirms that the interface that delimits the static inner cone has an approximately constant slope, regardless of the initial aspect ratio. Since the immobile region in regime~II forms an inner cone of base radius $R_0$ and height $H_\infty$, the near-constancy of its interface angle implies $\tan\theta_2^{\mathrm{II}}=H_\infty/R_0\simeq \mathrm{const.}$, providing a direct geometric interpretation of the height plateau value $C$ observed in figure~\ref{fig:Quatuor_depot}(a) and defined in \eqref{eq:Hf_regimeII}, so that $C = \tan \theta_2^{\mathrm{II}}$ . The outer angle $\theta_1$ also seems constant in regime~I with a smaller value $\theta_1^{\mathrm{I}} = (19.3 \pm 1.5)^\circ$.
{This value of $\theta_1^{\mathrm{I}}$ is expected to be slightly smaller than the value of angle of repose $\theta_r = (23.9 \pm 0.7)^\circ$ measured in the quasi-static flow deposit by the slow removal of the gate as described in section 2. Indeed, the angle of repose from an avalanche flow has been shown to be smaller for larger starting angle by \cite{FischerPRE2008}, and the starting angle imposed by the collapse experiments is much larger than the maximum angle of stability of the pile.}
In regime II, the outer angle $\theta_1$ is no longer constant but decreases when $a$ increases, from $\theta_1^{\mathrm{II}} \simeq 20^\circ$ at $a_c \simeq 0.7$ down to $\theta_1^{\mathrm{II}} \simeq 5^\circ$ at $a \simeq 10$. This decrease reflects the progressive flattening of the outer free surface as the aspect ratio increases: taller columns undergo a more inertial collapse, leading to a larger spreading $R_\infty$ while $H_\infty$ remains bounded, so that $H_\infty/R_\infty$ (and thus $\theta_1^{\mathrm{II}}$) decreases.

Altogether, figures~\ref{fig:Mg} and \ref{fig:Modele}(a) establish a robust geometric structure for the deposits in regimes~I and II: a static inner cone with a nearly constant slope $\theta_2$, surrounded by a flowing shell whose outer slope $\theta_1$ depends on the aspect ratio. In the next subsection, we translate this qualitative picture into a simple geometric model that will be used to rationalise the scaling laws of \S\ref{sec:res_scaling}.

\begin{table}
    \centering
    \begin{tabular}{c c c c c c}
    Angle       & Symbol                                 & Expression                                    & Mean value (°)    & Standard deviation (°)    \\ \hline 
      $\theta_1^\mathrm{I}$   & \raisebox{-0.2ex}{\textcolor{BlueViolet}{$\squadfill$}} & $\text{atan}\left( \frac{H_0}{R_\infty-R_c}\right)$& 19.3              & 1.5                       \\
      $\theta_2^\mathrm{I}$  & \large\textcolor{BrickRed}{$\circletfill$}  & $\text{atan}\left( \frac{H_0}{R_0-R_c}\right)$  & 35.6              & 3.0                       \\
      $\theta_1^\mathrm{II}$   & \textcolor{ProcessBlue}{$\blacktriangle$}      & $\text{atan}\left( \frac{H_\infty}{R_\infty}\right)$      & --                 & --                        \\
    $\theta_2^\mathrm{II}$    & \textcolor{Orange}{$\bigstar$}   & $\text{atan}\left( \frac{H_\infty}{R_0}\right)$      & 38.2              & 2.8                       \\

    \end{tabular}
    \caption{Expression of the angles $\theta_2$ and $\theta_1$ used for the measurements shown in figure~\ref{fig:Modele}(b).}
    \label{tab:theta}
\end{table}

\subsection{Geometric model for the deposit shape}
\label{sec:geom_model}

\newcommand{\Xc}{\frac{R_c}{R_0}}
\newcommand{\Xf}{\frac{R_\infty}{R_0}}
\newcommand{\A}{\frac{a}{\tan \theta_r}}

Given the observation made in the previous section, we will consider that the internal angle $\theta_2=\theta_2^\mathrm{I}=\theta_2^\mathrm{II}$ is a constant independent of the aspect ratio $a$ in the modelling developed below. 
For small aspect ratios, since $\tan \theta_2^\mathrm{I} = H_0/(R_0-R_c)$, the truncation radius is thus directly given by 

\begin{equation}
\frac{R_c}{R_0}
=
1-\frac{a}{\tan\theta_2^{\mathrm{I}}}. \label{eq:Rc}
\end{equation}

As the aspect ratio increases, $R_c$ is thus expected to decrease down to zero at the transition between regimes I and II, corresponding to the critical aspect ratio $a_c$. This gives a very simple relation between the critical aspect ratio $a_c$ and the angle $\theta_2$:

\begin{equation}
a_c = \tan\theta_2^{\mathrm{I}}. \label{eq:transi}
\end{equation}

The $R_c$ measurements in regime I are well described by the linear decrease predicted by (\ref{eq:Rc}), as shown in the inset of figure~\ref{fig:Quatuor_depot}(b), yielding $\tan \theta_2^{\mathrm{I}} = 0.7$, corresponding to the observed transition value $a_c = 0.7 \pm 0.1$. { Related stability models have also related granular or soil-column failure to a characteristic failure geometry and friction angle \citep{Morse2014,TaylorNoonan2021}. The present relation \eqref{eq:transi} has a different status: $a_c$ is identified geometrically from the disappearance of the truncated static core, while {$\theta_2^{\mathrm{I}}$ is inferred from the measured deposit geometry through the relations given in table~\ref{tab:theta}}. Equation~\eqref{eq:transi} is therefore used here as a geometrical closure rather than as a constitutive failure criterion.}

In regime I, enforcing volume conservation between the initial (quarter-)cylinder and the final (quarter-)truncated cone yields
\begin{equation}
{R_\infty^{\mathrm{I}}}^2+{R_\infty^{\mathrm{I}}} R_c + R_c^2 = 3R_0^2.
\label{eq:VolCons}
\end{equation}

Using (\ref{eq:Rc}) with $\tan\theta_2^{\mathrm{I}} = a_c$, the solution of the second-degree equation (\ref{eq:VolCons}) in $R_\infty^{\mathrm{I}}$ is

\begin{equation}
\frac{R_\infty^{\mathrm{I}}}{R_0}
=\frac{\sqrt{12-3\left( 1-a/a_c\right)^2} -\left( 1-a/a_c\right)}{2}  \quad  \text{for } a \leqslant a_c.
\label{eq:Rf_regimeI}
\end{equation}

The above equation captures the curvature of the regime-I runout data shown in figure \ref{fig:Quatuor_depot}(b), but the best fit gives the value $a_c^{\rm fit}=0.52$, which is significantly smaller than the transition value $a_c=0.7\pm0.1$ inferred from the disappearance of the truncation radius and from the static-core angle. In the following, we will use $a_c=0.7$ as the geometric value in the mass and energy model, while regarding the smaller fitted value for the runout as a measure of the limitations of the straight truncated-cone closure. Equation \eqref{eq:Rf_regimeI} predicts also that $R_\infty/R_0=\sqrt{3}$ when the truncation radius vanishes at $a=a_c$, in reasonable agreement with the experimental measurements near the transition between regimes~I and~II, as shown in figure \ref{fig:Quatuor_depot}(b).

Since $\tan \theta_1^\mathrm{I}=H_0/(R_\infty^\mathrm{I}-R_c)$, equation (\ref{eq:Rf_regimeI}) provides the following prediction for the pile angle $\theta_1^\mathrm{I}$ in regime I:
\begin{equation}
    \tan \theta_1^\mathrm{I} =\frac{(2a/a_c)\tan\theta_2}{\sqrt{12-3\left( 1-a/a_c\right)^2}  -3\left( 1-a/a_c\right)}  \quad  \text{for } a \leqslant a_c.
    \label{eq:theta_1_I}
\end{equation}

For vanishing aspect ratio ($a/a_c \ll 1$), equation (\ref{eq:theta_1_I}) reduces to $\tan \theta_1^\mathrm{I} \simeq \tan \theta_2 /2 \simeq a_c/2$, and at the transition between regimes I and II for $a/a_c = 1$, (\ref{eq:theta_1_I}) predicts that $\tan \theta_1^\mathrm{I} \simeq \tan \theta_2 /\sqrt{3} \simeq a_c/\sqrt{3}$, which is only a little larger than $\tan \theta_2 /2 \simeq a_c/2$. The full prediction of equation (\ref{eq:theta_1_I}) as a function of $a$ is plotted as a solid line in figure~\ref{fig:Modele}(b) using the model relation $a_c=\tan\theta_2$, with $a_c=0.7$, which is in rather good agreement with the data even if the slight increase of $\theta_1$ with the aspect ratio $a$ predicted by equation (\ref{eq:theta_1_I}) in regime I is not observed in the experimental data.
{With $\tan\theta_2 =0.7$, this leads to a predicted $\tan\theta_1^\mathrm{I} $ in regime I between $\tan\theta_2/2 =0.35$ and $\tan\theta_2/\sqrt{3} \simeq 0.4$, and thus to a predicted $\theta_1^\mathrm{I}$ between $19.3^\circ$ and $21.8^\circ$, which corresponds well to the experimentally measured value $\theta_1^\mathrm{I} = (19.3 \pm 1.5)^\circ$.}
{As already discussed in section~\ref{sec:angles}, the angle $\theta_1^\mathrm{I}$ is slightly smaller than the angle of repose $\theta_r$. Since the geometric model gives $\tan\theta_1^\mathrm{I} \simeq \tan\theta_2/2$, this is consistent with $\tan\theta_2 \simeq 0.7$ being slightly smaller than $2\tan\theta_r \simeq 0.9$.}

{The measured angle $\theta_2$ itself is the slope of the static--mobile interface and should therefore be identified neither with the angle of repose nor with the material friction angle. Its possible connection with an internal failure angle \citep{Lajeunesse2004Spreading} is an interpretation of the observed geometry rather than an independently imposed constitutive property, and its numerical value could be material- and preparation-dependent and not a universal constant.}

In regime II, where the deposit may be approximated by a cone when neglecting the slight curvature of the free surface, volume conservation between the initial (quarter-)cylinder, $V_0=\pi R_0^2H_0/4$, and the final (quarter-)cone, $V_f=\pi {R_\infty^{\mathrm{II}}}^2 H_\infty^{\mathrm{II}}/12$, gives the simple relation
\begin{equation}
3H_0{R_0}^2 = H_\infty^{\mathrm{II}}{R_\infty^{\mathrm{II}}}^2.
\label{eq:VolConsII}
\end{equation}
By using $H_\infty^{\mathrm{II}}/R_0 = \tan \theta_2^{\mathrm{II}} =C \simeq a_c$ from \eqref{eq:Hf_regimeII}, the conservation relation \eqref{eq:VolConsII} yields
\begin{equation}
\frac{R_\infty^{\mathrm{II}}}{R_0} \simeq \left(\frac{3a}{a_c}\right)^{1/2}  \quad  \text{for } a \geqslant a_c.
\label{eq:Rf_regimeII}
\end{equation}
The above model equation, already introduced by \cite{Lajeunesse2004Spreading}, implies that the numerical prefactor $D$ of \eqref{eq:Rf_scaling} should be $D_{\rm model} \simeq (3/a_c)^{1/2} \simeq (3/\tan\theta_2)^{1/2}$.
Using $a_c=0.7\pm0.1$ gives $D_{\rm model}=2.1\pm0.2$, which is a little smaller than the fitted value $D=2.40\pm0.03$ in figure~\ref{fig:Quatuor_depot}(b). {Using the two measured mean values of $\theta_2$ separately changes
$D_{\rm model}$ by less than $5\,\%$. The slight underprediction of $D$ likely reflects the departure of the regime-II deposit from a perfect cone or from a bulk solid fraction at the final state $\phi_\infty$ different from the initial bulk solid fraction $\phi_0$. For our experiments the ratio $\phi_\infty/\phi_0$ is on average 0.99 with a standard deviation of 0.11. But a 10\% difference between $\phi_\infty$ and $\phi_0$ would modify $D_{\rm model}$ by only 5\%.
Within the
small-$a$ relation $\tan\theta_2\simeq 2\tan\theta_r$, the prefactor can also be written as $D_{\rm model}\simeq [3/(2\tan\theta_r)]^{1/2}$. }

Concerning the angle of the final deposit in regime II, which is given by $\tan\theta_1^{\mathrm{II}} = H_\infty^{\mathrm{II}}/R_\infty^{\mathrm{II}}$, the combination of  \eqref{eq:Hf_regimeII}, with $C \simeq \tan\theta_2$, and \eqref{eq:Rf_regimeII} gives

\begin{equation}
\tan\theta_1^{\mathrm{II}}
= \tan\theta_2
\left(\frac{a_c}{3a}\right)^{1/2}  \quad  \text{for } a \geqslant a_c.
 \label{eq:theta_1_II}
\end{equation}

Equation \eqref{eq:theta_1_II} predicts that $\tan\theta_1^{\mathrm{II}}$ should decrease as $a^{-1/2}$ in regime II from the value $\tan \theta_2 /\sqrt{3}$ at $a=a_c$. The prediction by \eqref{eq:theta_1_II}, with $a_c=\tan\theta_2=0.7$, is in rather good agreement with the experimental data of regime II as shown by the solid line in figure~\ref{fig:Modele}(b).

This geometric closure will be used in section \ref{sec:mass_energy} to derive the partition of mass of flowing grains relative to static grains
and to interpret the final potential-energy loss of the collapse.

\section{Partition of mass and final energy budget}
\label{sec:mass_energy}

\subsection{Mass partition: exiting and flowing masses}
\label{section:mass-partition}

\begin{figure}
    \centering
    \includegraphics[width=\linewidth]{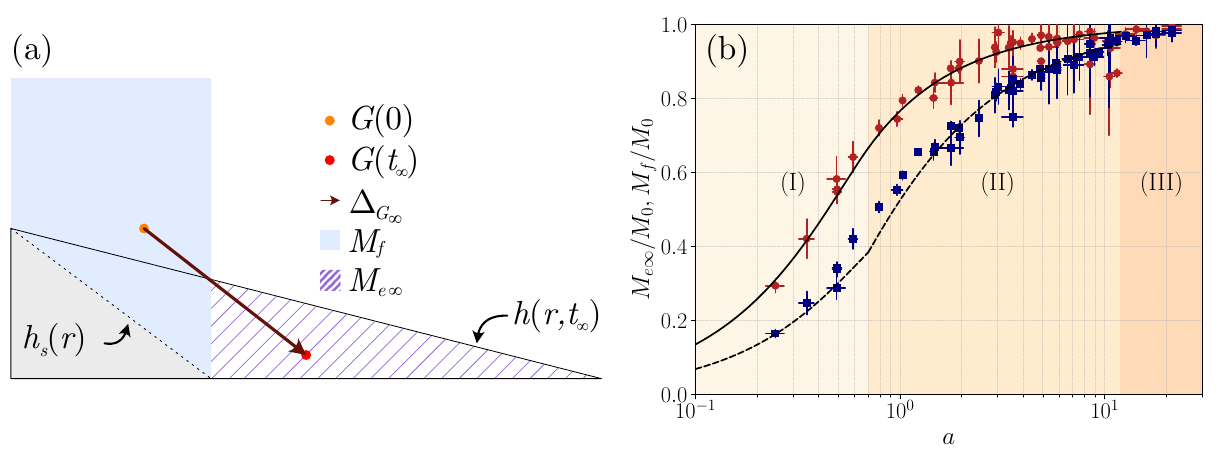}
    \caption{(a) Schematic definition of the flowing mass $M_f$ (blue region), exiting mass $M_e$ (hatched region) and static mass $M_s = M_0 - M_f$ (grey region). (b) Flowing-mass fraction $M_f/M_0$ (\raisebox{-0.2ex}{\Large\textcolor{BrickRed}{$\bullet$}}) and exiting-mass fraction $M_e/M_0$ (\textcolor{BlueViolet}{$\blacksquare$}) as functions of the aspect ratio $a$. Error bars are not shown because estimating the corresponding uncertainties is difficult; the scatter provides an order-of-magnitude estimate. The solid lines correspond to the {model relations} for $M_f/M_0$ (equations~\ref{eq:MfI_a} for $a<a_c$ and \ref{eq:MfII} for $a_c<a<12$ with $a_c=0.7$), and the dashed lines to the {model relations} for $M_e/M_0$ (equations~\ref{eq:MsI} for $a<a_c$ and \ref{eq:MsII} for $a_c<a<12$ with $a_c=0.7$). }
    \label{fig:schema_Ms_Mg}
\end{figure}

In their analysis of the collapse of a granular column, \cite{Staron2005} focused on the mass of grains that leave the volume initially occupied by the column. This exiting mass, denoted here $M_e$, corresponds to the hatched region in the schematic of figure \ref{fig:schema_Ms_Mg}(a). All the grains of this region have flowed, but the grains that have flowed are not all in this region. From the image-differencing maps of figure~\ref{fig:Mg}, we know that all the flowing grains are located in the portion of the initial column above the static region, which appears in blue in the schematic of figure \ref{fig:schema_Ms_Mg}(a). We here define the \emph{flowing} mass $M_f$ as the mass of grains that move at least once during the slump. By definition, $M_e\leq M_f$, since some grains may move without exiting the initial footprint.

We estimate both $M_e$ and $M_f$ from binarised images using difference operations. Let $A_t(t)$ be the total visible area of grains at time $t$, and let $r_{A_t}(t)$ be the radial coordinate of its planar centroid in the meridional $(r,z)$ plane. Subtracting the binarised initial frame from the frame at time $t$ isolates the grains that have left the initial footprint, and we denote the corresponding area by $A_e(t)$ and the radial coordinate of its planar centroid by $r_{A_e}(t)$. Likewise, the region occupied by grains that have moved can be reconstructed by intersecting, at each time $t$, the binary mask of the whole granular assembly with the binary mask of pixels flagged as \emph{moving} from frame-to-frame differencing (figure~\ref{fig:Mg}).  Using Guldinus' theorem to convert areas into volumes from axisymmetry, one can write
\begin{equation}
V_t(t)= \frac{\pi}{2} r_{A_t}(t)\,A_t(t), 
\qquad
V_e(t)= \frac{\pi}{2} r_{A_e}(t)\,A_e(t),
\qquad
V_f(t)= \frac{\pi}{2} r_{A_f}(t)\,A_f(t),
\label{eq:MsMf_t}
\end{equation}
for the instantaneous total volume of grains, $V_t$, volume $V_e$ of exiting grains and volume $V_f$ of flowing grains, respectively. Here, $r_{A_t}$, $r_{A_e}$ and $r_{A_f}$ denote the radial coordinates of planar centroids used only for the area-to-volume conversion. 
By considering mass conservation $M(t) = M_0$ along the entire collapse and a homogeneous solid fraction $\phi(t)$ at each time $t$ among all the granular assembly, the exiting and flowing masses $M_e(t)$ and $M_f(t)$can be deduced from $V_e(t)$, $V_f(t)$, $V_t(t)$ and $M_0$ as follows:
\begin{equation}
M_e(t)= \frac{V_e(t)}{V_t(t)}\,M_0, 
\qquad
M_f(t)= \frac{V_f(t)}{V_t(t)}\,M_0.
\label{eq:MsMf_t_2}
\end{equation}

{ Indeed, if the instantaneous solid fraction $\phi(t)$ is considered to be the same for all the instantaneous volumes $V_e(t)$, $V_f(t)$, $V_t(t)$, it cancels from the ratio $V_e(t)/V_t(t)$ and $V_f(t)/V_t(t)$ of partial to total volumes at any given time $t$.} The exiting mass $M_e$ is expected to increase from zero to a final value $M_{e \infty}=M_e(t_\infty)$ once the deposit has ceased to evolve and this is what is observed. By contrast, the flowing mass $M_f$ is expected to be constant, and is indeed found to be quasi-constant during the collapse.
In practice, the total flowing area $A_f$ is obtained by identifying the grains that have moved at least once over the whole sequence (sum of moving pixels). This yields an instantaneous area $A_f(t)$, whose planar centroid has radial coordinate $r_{A_f}(t)$.
{We checked that varying slightly the binarisation threshold around the chosen value does not change significantly the value obtained for $M_f$.}

Figure~\ref{fig:schema_Ms_Mg}(b) reports the measured fractions $M_{f}/M_0$ and $M_{e \infty}/M_0$ as a function of the aspect ratio $a$. Both $M_{f}/M_0$ and $M_{e \infty}/M_0$ increase monotonically with $a$, and both tend toward $1$ at very large aspect ratios. As expected, the exiting mass $M_e$ is always smaller than the flowing mass $M_f$.

Using the simplified geometry introduced in \S\ref{sec:geom_model} together with the deposit-size scalings of \S\ref{sec:res_scaling}, we can derive analytical expressions for the flowing and exiting masses, $M_{f}$ and $M_{e \infty}$ in regimes I and II as detailed in appendices \ref{appendix:regimeI} and \ref{appendix:regimeII}. 

In regime~I, the flowing mass corresponds to the grains located above the inner stagnant interface in the region $R_c<r\leq R_0$. Using \eqref{eq:Rc} and \eqref{eq:transi} for $R_c$ (see appendix \ref{appendix:regimeI_flowing} for details) gives

\begin{equation}
\frac{M_{f}^{\rm I}}{M_0}
= \frac{a}{a_c}
\left(
1-\frac{a}{3 a_c}
\right)  \quad  \text{for } a \leqslant a_c.
\label{eq:MfI_a}
\end{equation}

In regime~II, the flowing mass is simply the difference between the initial quarter-cylinder and the stagnant inner cone of height $H_\infty$. Using the plateau law $H_\infty/R_0=C\simeq a_c$ (see appendix \ref{appendix:regimeII_flowing} for details) gives
\begin{equation}
\frac{M_{f}^{\mathrm{II}}}{M_0} \simeq 1-\frac{a_c}{3a}  \quad  \text{for } a \geqslant a_c.
\label{eq:MfII}
\end{equation}

We now turn to the exiting mass $M_e$, defined as the mass of grains that leave the initial footprint of the column. In regime~I, this corresponds to the portion of the truncated-cone deposit located at $r>R_0$.
Using \eqref{eq:Rc} and \eqref{eq:transi} for $R_c$ and \eqref{eq:Rf_regimeI} for $R_\infty^{\rm I}$ (see appendix \ref{appendix:regimeI_exiting} for details) leads to

\begin{equation}
\frac{M_{e \infty}^{\rm I}}{M_0}
= \frac{2}{3}
\frac{
 \left(3-a/a_c\right) (a/a_c)^2
}{\sqrt{12-3\left( 1-a/a_c\right)^2}  -3\left( 1-a/a_c\right)}  \quad  \text{for } a \leqslant a_c.
\label{eq:MsI}
\end{equation}

In regime~II, the grains remaining within the initial footprint occupy the intersection of the final cone with the initial cylinder $r\leq R_0$. Using \eqref{eq:Hf_regimeII} with $C \simeq a_c$ for $H_\infty^{\rm II}$ together with \eqref{eq:Rf_regimeII} for $R_\infty^{\rm II}$ (see appendix \ref{appendix:regimeII_exiting} for details) leads to 

\begin{equation}
\frac{M_{e \infty}^{\mathrm{II}}}{M_0}
= 1-\frac{a_c}{a}+\frac{2}{3\sqrt{3}}\,\left(\frac{a_c}{a}\right)^{3/2}  \quad  \text{for } a \geqslant a_c.
\label{eq:MsII}
\end{equation}

{With $a_c$ determined independently from the deposit geometry in \S\ref{sec:geom_model}, no additional parameter is fitted to the flowing- or exiting-mass data.}

{The resulting relations for the flowing- and exiting-mass fractions} obtained from the geometric model equations~\eqref{eq:MfI_a}--\eqref{eq:MfII} and \eqref{eq:MsI}--\eqref{eq:MsII} are plotted in figure~\ref{fig:schema_Ms_Mg}(b) as a function of the aspect ratio $a$ (solid lines for $M_{f}/M_0$ and dashed lines for $M_{e \infty}/M_0$). Overall, the agreement with the measurements is good over most of the explored range for both mass fractions of interest, $M_{f}/M_0$ and $M_{e \infty}/M_0$. Importantly, the difference between flowing and exiting masses decreases rapidly as $a$ increases: for $a\gtrsim 3$ the relative difference $(M_{f}-M_{e \infty})/M_{e \infty}$ is typically below $10\%$, and it drops below $5\%$ for $a\gtrsim 8.5$. In this moderate-to-large aspect-ratio range, the exiting mass can therefore be used as a practical estimate of the flowing mass, as in, e.g., \citealp{Staron2005} whose ``sliding mass'' denoted $M_s$ corresponds to our exiting mass denoted $M_e$. Altogether, figure~\ref{fig:Mg} and the trends in figure~\ref{fig:schema_Ms_Mg}(b) indicate that the mass partition is primarily governed by the aspect ratio: at small $a$ only a thin peripheral annulus is mobilised, whereas at large $a$ most of the grains participate in the motion. This partition will be used below to interpret the final potential-energy loss of the collapse in terms of an effective frictional work.

\subsection{Barycentric displacements}
\label{sec:Barycentre}

From the instantaneous measured interface $h(r,t)$, the instantaneous barycentre $G(t)=(r_G(t),z_G(t))$ of the granular material contained in any slice of elementary azimuthal extension ${\rm d}\theta$ can be extracted. %
{ Assuming a homogeneous packing fraction within the slice, the radial and vertical coordinates of $G(t)$ are obtained from the corresponding volume moments, whose explicit expressions are given in appendix~\ref{app:barycentre_allgrains}.}

{Here $r_G$ denotes the radial coordinate of the barycentre of an elementary azimuthal slice and should not be confused with the planar centroid $r_{A_t}$ introduced in \S\ref{section:mass-partition}, which is used only for the area-to-volume conversion. Considering an azimuthal slice preserves information on the radial redistribution of the grains, which would be lost when considering a full cylindrical geometry, whose barycentre remains on the symmetry axis. The slice barycentre $G$ therefore provides a compact measure of both the radial spreading and the vertical fall of the granular mass.
An example of its motion during the collapse is shown by the successive black dots in figure~\ref{fig:effondrements}(a).} The instantaneous displacement $\Delta_G(t)$ from the initial barycentre $G(0)$ is defined as
\begin{equation}
\Delta_G(t) = \sqrt{ \left[r_G(t)-r_G(0)\right]^2 + \left[z_G(t)-z_G(0)\right]^2 }.
\label{eq:DeltaG}
\end{equation}
The quantity $\Delta_G(t)$ corresponds to the net displacement between the initial position $G(0)$ and the current position $G(t)$ of the barycentre, and does not correspond exactly to the length of the real curved path followed by the barycentre, but is a little smaller. In the following, we denote by $\Delta_{G\infty}=\Delta_G(t_\infty)$ its final value.

\begin{figure}
    \centering
    \begin{subfigure}[b]{0.45\textwidth}
         \centering
         \includegraphics[width=\textwidth]{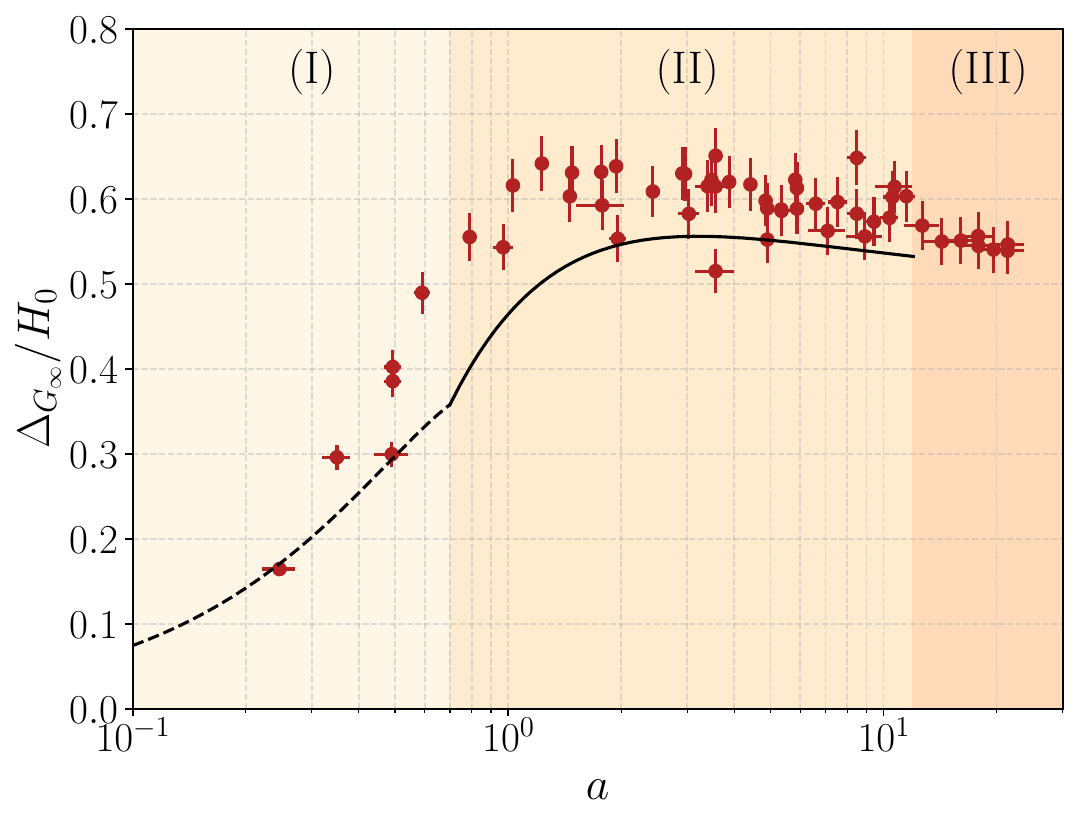}
         \caption{}
         \label{fig:barycentre_dist}
     \end{subfigure}
     \hfill
     \begin{subfigure}[b]{0.45\textwidth}
         \centering
         \includegraphics[width=\textwidth]{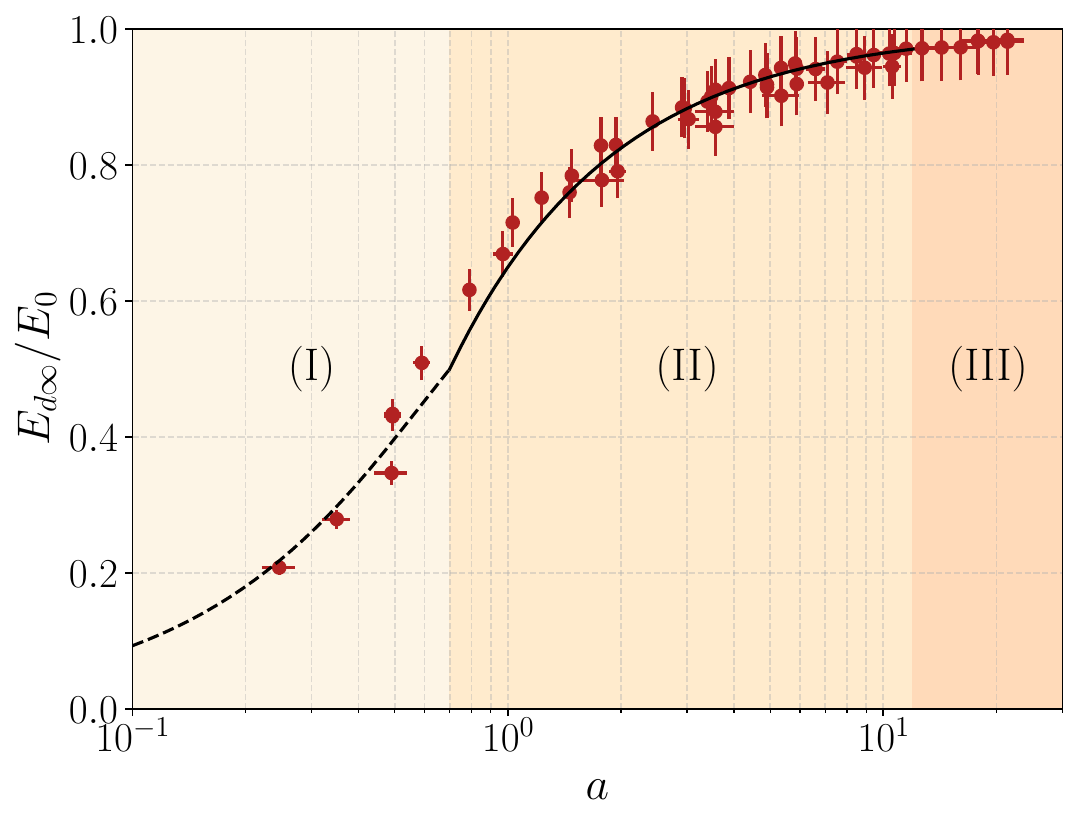}
         \caption{}
         \label{fig:NRJs}
     \end{subfigure}
     \hfill
   \caption{(a) Normalised final barycentre displacement $\Delta_{G \infty}/H_0$ and (b) dissipated-energy fraction $E_{d \infty}/E_0$ as functions of the aspect ratio $a$. Red symbols correspond to the experimental measurements. { In panel~(a), the dashed line follows from equation~\eqref{eq:baryI}, while the solid line corresponds to equation~\eqref{eq:DeltaGII}. In panel~(b), the dashed and solid lines correspond to equations~\eqref{eq:EdissI} and \eqref{eq:EdissII}, respectively. All model curves use $a_c=0.7$.}} %
    \label{fig:barys}
\end{figure}

Figure~\ref{fig:barys}(a) reports $\Delta_{G\infty}/H_0$ as a function of the aspect ratio $a$. The normalised displacement is small at low $a$, increases rapidly across regime~I and the beginning of regime~II, and reaches a plateau of approximately $0.6$ at $a \gtrsim 1$ before decreasing slightly toward approximately 0.5 at larger aspect ratios. Within the simplified description of the final deposit as a truncated cone in regime~I or a full cone in regime~II, the final position $G_\infty(r_{G\infty},z_{G\infty})$ of the slice barycentre can be obtained analytically, and then the net displacement $\Delta_{G\infty}$ of the slice barycentre from its initial position $G(0)$ at $(r_G(0) =2R_0/3, z_G(0)= H_0/2)$ corresponding to the initial column of height $H_0$ and radius $R_0$. The detailed analytical calculations can be found in appendix \ref{app:barycentre_allgrains} but we recall below the final expressions.

In regime~II where the final deposit is approximated as a cone of height $H_\infty$ and base radius $R_\infty$, the slice barycentre $G_\infty$ is located at
$(r_{G\infty}^{\mathrm{II}},z_{G\infty}^{\mathrm{II}})=(R_\infty/2,H_\infty/4)$ (see appendix \ref{app:barycentre_allgrains} for details). Using \eqref{eq:Hf_regimeII} with $C \simeq a_c$ for $H_\infty$ and \eqref{eq:Rf_regimeII} for $R_\infty$ gives the final normalised positions
\begin{equation}
\frac{z_{G\infty}^{\mathrm{II}}}{H_0} \simeq \frac{a_c}{4a} 
\quad
\text{and}
\quad
 \frac{r_{G\infty}^{\mathrm{II}}}{R_0} = \frac{1}{2}\left(\frac{3a}{a_c}\right)^{1/2} \quad  \text{for } a \geqslant a_c.
\label{eq:baryII}
\end{equation}
The corresponding final net displacement of the slice barycentre from its initial position is therefore
\begin{equation}
\frac{\Delta_{G\infty}^{\mathrm{II}}}{H_0} \simeq \left[ \frac{1}{a^{2}} \left( \left(\frac{3a}{4a_c}\right)^{1/2}-\frac{2}{3} \right)^2 + \left(\frac{1}{2}-\frac{a_c}{4a} \right)^2 \right]^{1/2}  \quad  \text{for } a \geqslant a_c.
\label{eq:DeltaGII}
\end{equation}
At large aspect ratios ($a \gg a_c$), the radial contribution to $\Delta_{G\infty}/H_0$ becomes small, so that the barycentric displacement is increasingly dominated by the vertical fall and approaches $\Delta_{G\infty}^{\mathrm{II}}/H_0=1/2$.

In regime~I, the calculation detailed in appendix~\ref{app:barycentre_allgrains} gives the final normalised positions of the slice barycentre $G$ as
\begin{equation}
\begin{aligned}
\frac{z_{G\infty}^{\mathrm{I}}}{H_0} &= \frac{ 6+3\left(1-a/a_c\right)^2+\left(1-a/a_c\right) \sqrt{12-3\left(1-a/a_c\right)^2}}{24} \quad  \text{for } a \leqslant a_c,
\\
\frac{r_{G\infty}^{\mathrm{I}}}{R_0} &= \frac{ 3\sqrt{12-3\left(1-a/a_c\right)^2}-3\left(1-a/a_c\right) +2\left(1-a/a_c\right)^3 }{12} \quad  \text{for } a \leqslant a_c.
\end{aligned}
\label{eq:baryI}
\end{equation}
The model prediction for $\Delta_{G\infty}^{\mathrm{I}}$ follows directly by substituting \eqref{eq:baryI} into \eqref{eq:DeltaG}.
In the limit $a\to 0$, equation~\eqref{eq:baryI} recovers the initial position $G(0)=(2R_0/3,H_0/2)$, and therefore
$\Delta_{G\infty}^{\mathrm{I}}\to0$. At the transition $a=a_c$, it gives
$z_{G\infty}^{\mathrm{I}}/H_0=1/4$ and $r_{G\infty}^{\mathrm{I}}/R_0=\sqrt{3}/2$
corresponding to the barycentre of a cone of radius
$R_\infty=\sqrt{3}R_0$. These values coincide with the regime-II equations \eqref{eq:baryII} and \eqref{eq:DeltaGII} at $a=a_c$.

The predictions for regimes~I and II by \eqref{eq:baryI} and \eqref{eq:DeltaGII} with $a_c = 0.7$ are shown in
figure~\ref{fig:barys}(a) by the dashed  and solid lines, respectively: $\Delta_{G\infty}/H_0$ increases with $a$ in regime I and at the beginning of regime II before reaching a maximum at $a \simeq 3$ and then slightly decreasing toward 1/2. These model predictions capture the overall evolution of the experimental measurements well.

\subsection{Energy loss and effective friction coefficients} \label{sec:energetic_dissipation}

Since the grains are at rest in both the initial and final states, the total energy $E_{d\infty}$ dissipated during the collapse can be obtained directly from the loss of gravitational potential energy $E_0-M_0 g z_G$ from its initial value $E_0 = M_0 g H_0/2$, so that the fraction of dissipated energy is given by
\begin{equation}
\frac{E_{d\infty}}{E_0} = 1-2\frac{z_{G\infty}}{H_0}.
\label{eq:Ediss_general}
\end{equation}

This fraction of dissipated energy can be calculated from either the experimental measurements of $z_{G \infty}$ or its model prediction using the expression given by \eqref{eq:baryII} and \eqref{eq:baryI}. The evolution of the fraction of dissipated-energy 
$E_{d\infty}/E_0$ with the aspect ratio $a$ is reported in figure~\ref{fig:barys}(b). The experimental data show that this fraction increases with $a$ from approximately 0.2 at $a \simeq 0.2$ up to approximately 1 at large $a$, showing that most of the initial potential energy is dissipated for columns with a large aspect ratio. Indeed, more than $90\%$ of the initial potential energy is dissipated for $a\gtrsim4$, and this fraction exceeds $95\%$ for $a\gtrsim8$. For small aspect ratios, only a limited portion of the column is mobilised, and the final potential-energy loss remains relatively small. As $a$ increases, a larger fraction of the grains participates in the collapse, and the final barycentre approaches the substrate, so that an increasing fraction of the initial potential energy is dissipated. At the transition between regimes I and II corresponding to $a \simeq 0.7$, the dissipated energy is approximately half the initial energy.
Substituting the final vertical position of the slice barycentre obtained in regimes~I and II and given in \eqref{eq:baryII} and \eqref{eq:baryI} into
\eqref{eq:Ediss_general}, and using the corresponding deposit-size relations, yields

\begin{align}
\frac{E_{d\infty}^{\mathrm{I}}}{E_0} &=
\frac{6 -3\left(1-a/a_c\right)^2 -\left(1-a/a_c\right)
\sqrt{12-3\left(1-a/a_c\right)^2}}{12} \quad  \text{for } a \leqslant a_c,
\label{eq:EdissI}
\\
\frac{E_{d\infty}^{\mathrm{II}}}{E_0} &=
1-\frac{a_c}{2a} \quad  \text{for } a \geqslant a_c.
\label{eq:EdissII}
\end{align}

The model predicts that the fraction of dissipated energy $E_{d\infty}/E_0$ increases from 0 at vanishing $a$ up to 1 for very large $a$, as expected, with the remarkable value 1/2 at the transition from regime I to regime II for $a/a_c = 1$. More precisely, the increase of $E_{d\infty}/E_0$ with $a$ is  first linear, $E_{d\infty}/E_0 \simeq (2/3)a/a_c$ for $a/a_c \ll 1$, and then non-linear before approaching the asymptotic value 1 for $a/a_c \gg 1$ with a difference decreasing as $a_c/2a$. All the model predictions by \eqref{eq:EdissI} and \eqref{eq:EdissII} shown in dashed and solid lines in figure~\ref{fig:barys}(b) for $a_c = 0.7$ capture the experimental data very well. { This global energy balance quantifies the total energy dissipated during the collapse, but does not resolve how this dissipation is partitioned between basal friction, interparticle friction, collisions and internal shear during the transient motion. This final-state energy balance is complementary to approaches that relate granular-column mobility to the transient kinetic energy developed during the collapse \citep{Topin2012,Polania2024}. Such approaches characterise the intermediate dynamics, whereas the present estimate follows directly from the difference between the initial and final static states.}

To interpret this final energy loss in terms of frictional work, we can consider the characteristic displacement $\Delta_{G\infty}$ of the slice barycentre with the total mass $M_0$, which leads to $E_{d \infty} = \mu_{\rm eff} \, M_0 \, g \, \Delta_{G_\infty}$ with an effective friction coefficient $\mu_{\rm eff}$. The parameter $\mu_{\rm eff}$ is not a local or instantaneous friction coefficient but a global estimate from the net work needed to bring the initial column to the final deposit. { In that sense, $\mu_{\rm eff}$ is analogous to global mobility estimates made for landslides \citep{Heim1932,LiedBakkehoi1980}, and {characterises} the integrated outcome of the motion rather than an instantaneous basal friction coefficient. This collapse-integrated viewpoint is complementary to previous energetic analyses: \citet{Staron2005} separated the energy transferred to the lateral flow and showed that its dissipation could be represented by an effective basal friction, while \citet{BalmforthKerswell2005} incorporated basal friction in a shallow-flow description of planar slumps.} The same estimate of $\mu_{\rm eff}$ can be made using only the flowing mass and the displacement $\Delta_{G_{f\infty}}$ of the slice barycentre of the flowing grains, which are located between $h(r,t)$ and $h_s(r)$ (see appendix \ref{app:barycentre_flowinggrains} for the exact definition). Both approaches  are in fact equivalent since the static part is unchanged between the initial and final states, so that the total and flowing-grain barycentric balances give  $M_0 \, \Delta_{G_\infty} = M_f \, \Delta_{G_{f\infty}}$. The flowing mass $M_f$ is smaller than the total mass $M_0$, but the characteristic displacement $\Delta_{G_{f\infty}}$ of the flowing grains is larger than the characteristic displacement $\Delta_{G_\infty}$ of all the grains within any elementary azimuthal slice. Thus, the collapse-integrated effective friction coefficient $\mu_{\rm eff}$ is given by
\begin{equation}
\mu_{\rm eff} = \frac{E_{d \infty}}{M_0 \, g \, \Delta_{G_\infty}} = \frac{E_{d \infty}}{M_f \, g \, \Delta_{G_{f\infty}}}.
\label{eq:mu_def}
\end{equation} 

\begin{figure}
    \centering
    \includegraphics[width=0.5\linewidth]{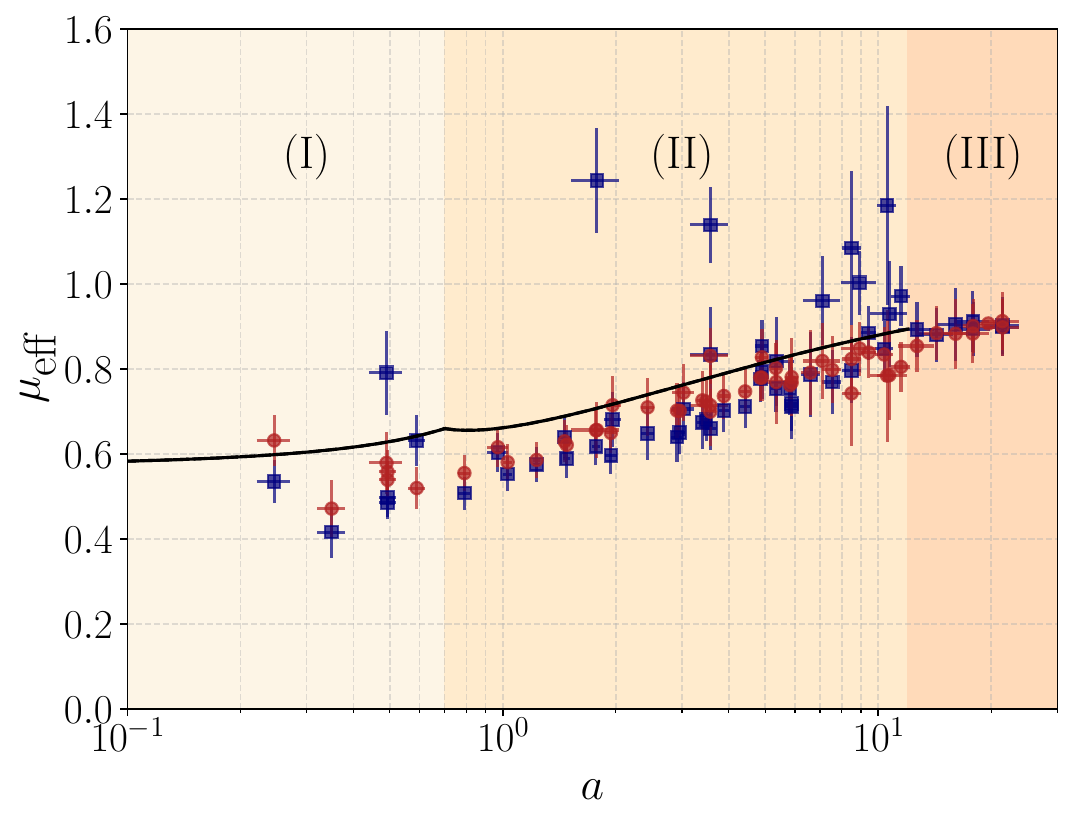}
    \caption{Effective friction coefficient $\mu_{\rm eff}$ as a function of the aspect ratio $a$. The data points are measurements from the ratio $E_{\mathrm{diss}}/(M_0 \, g \, \Delta_G)$ corresponding to all the grains (\textcolor{BrickRed}{$\circletfill$}) or from the ratio $E_{d \infty}/(M_f\, g \, \Delta_{G_{f\infty}})$ corresponding to only the flowing grains (\textcolor{BlueViolet}{$\circletfill$}). The solid line corresponds to the prediction of the model with $a_c = 0.7$.}
    \label{fig:Ediss_M}
\end{figure}

We report in figure~\ref{fig:Ediss_M} the two experimental estimates for $\mu_{\rm eff}$, either from $M_0 \Delta_{G_\infty}$ or from $M_f \Delta_{G_{f \infty}}$ as a function of the aspect ratio $a$. As expected, the estimate based only on the flowing grains is more scattered, but it follows the same trend as the estimate based on all the grains, increasing from approximately 0.5 at low $a$ to approximately 0.9 at large $a$.
The model prediction by \eqref{eq:mu_def}, with \eqref{eq:EdissI}, \eqref{eq:MfI_a} and \eqref{eq:baryI} in regime I, and with \eqref{eq:EdissII}, \eqref{eq:MfII} and \eqref{eq:DeltaGII} in regime II, for $a_c = 0.7$, is shown by the solid line in figure~\ref{fig:Ediss_M} and matches quite well the experimental measurements, with a significant increase of $\mu_{\rm eff}$ from approximately 0.6 at the transition from regime I to regime II up to approximately 0.9 close to the transition from regime II to regime III, but no significant variation of $\mu_{\rm eff}$ in regime I. 
{ The following subsection examines whether this evolution of $\mu_{\rm eff}$ can be parametrised using a macroscopic inertial number inspired by the functional form of the local $\mu(I)$ rheology.}

\subsection{{A macroscopic $\mu_{\rm eff}(I^*)$ parametrisation}}
\label{sec:muI}

\begin{figure}
    \centering
    \begin{subfigure}[b]{0.45\textwidth}
        \includegraphics[width=\textwidth]{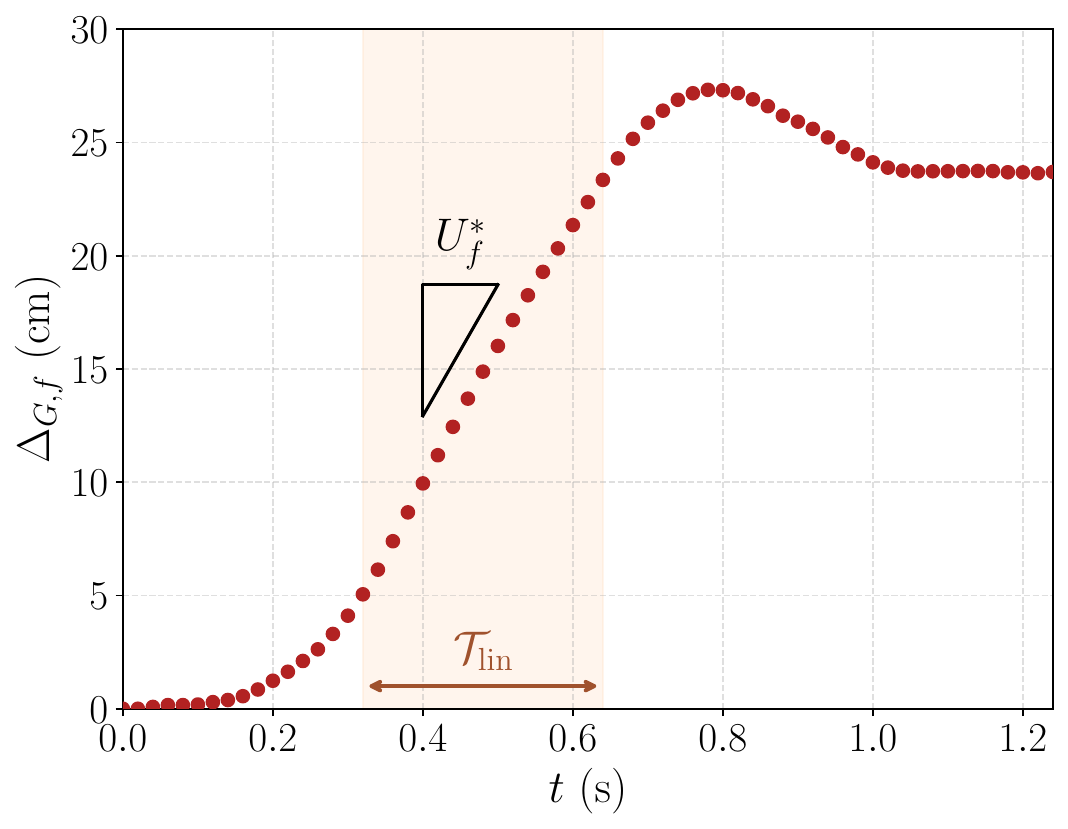}
        \caption{}
        \label{fig:Dg_t}
    \end{subfigure}
    \hfill
    \begin{subfigure}[b]{0.45\textwidth}
        \includegraphics[width=\textwidth]{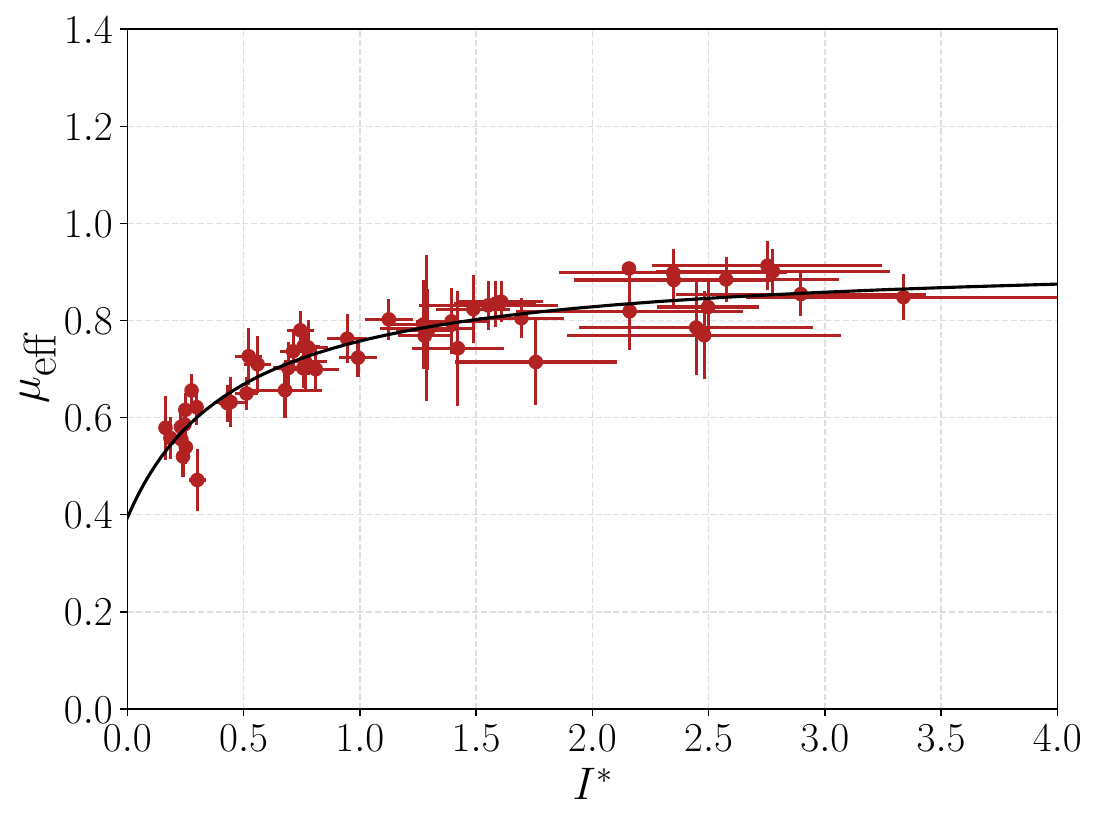}
        \caption{}
        \label{fig:mu_I}
    \end{subfigure}
    \hfill
    \caption{(a) Time evolution of $\Delta_{G,f}(t)$ for the same configuration as in figure \ref{fig:effondrements}(a) ($H_0=30$ cm, $R_0=20$ cm). (b) Evolution of $\mu_{\rm eff}$ with the macroscopic inertial number $I^*$. { The solid line corresponds to the best fit of the data by \eqref{eq:mu(I)} with $\mu_-=0.39$, $\mu_+=0.93$, and $I_0^*=0.48$.}}
\label{fig:Trio_friction}
\end{figure}

The local $\mu(I)$-rheology, which has been established for steady homogeneous flows \citep{Andreotti2013}, has also been demonstrated to be a useful framework for some unsteady and nonhomogeneous flows, such as the collapse of columns \citep{Lacaze2009,Lagree2011,Valette2019}. In this local rheology, {for a given granular material, the local friction coefficient $\mu$ is expressed as a function of the local inertial number
$I = \dot{\gamma} d/\sqrt{p/\rho}$ through a constitutive law involving material-dependent parameters}, where $\dot{\gamma}$ is the local shear rate, and $p$ is the local granular pressure. { In our experiments, neither $p$ nor $\dot{\gamma}$ is measured locally. The quantities introduced below are therefore characteristic bulk estimates for a strongly unsteady and spatially heterogeneous collapse, constructed from the structure and dynamics of the flowing grains.} Indeed, the grains belonging to the static core do not participate in the shear and should therefore not contribute to the characteristic flowing-layer thickness or pressure. To estimate the shear rate, we have to estimate both the characteristic shear velocity and the characteristic shear thickness. The characteristic velocity can be estimated from the barycentric motion of the flowing grains.
Figure~\ref{fig:Trio_friction}(a) shows a typical time evolution of the displacement of the barycentre $\Delta_{G_f}(t)=\|\boldsymbol{G_{f}(0)G_f(t)}\|$ corresponding to only the flowing grains for the typical collapse of figure \ref{fig:effondrements}(a). After an initial acceleration, this displacement exhibits an approximately linear increase during an intermediate stage of the collapse, before reaching a maximum and relaxing towards its final value. The characteristic velocity of the collapse can be determined from a linear fit over this intermediate interval, denoted by $\mathcal{T}_{\mathrm{lin}}$:
\begin{equation}
U_f^* = \left. \frac{{\rm d}\Delta_{G_f}}{{\rm d}t}
\right|_{\mathcal{T}_{\mathrm{lin}}}.
\label{eq:Ustar}
\end{equation}
The characteristic shear thickness $h^*$ can be estimated by the ratio $h^* = V_f/A^*$, where $V_f$ is the volume of flowing grains and {$A^*=\pi R_\infty^2/4$ is the characteristic horizontal footprint area of the final quarter-circular deposit. Thus, $h^*$ provides a bulk estimate of the mobilised-layer thickness obtained from the spreading of the volume $V_f$ over the basal footprint $A^*$.} This characteristic thickness scales remarkably with the final height $H_\infty$ in both regimes I and II: $h^* \simeq 0.2 H_\infty$, as shown in appendix~\ref{app:hstar}.

The characteristic shear rate can then be estimated as $\dot{\gamma}^*=U_f^*/h^*$ and the characteristic pressure at the base of the flowing layer as $P^* = \phi \rho g h^*$, so that the macroscopic inertial number $I^*$ characteristic of the collapse of a granular column can be estimated as
\begin{equation}
I^* = \frac{\dot{\gamma}^*d}{\sqrt{P^*/\rho}} =
\frac{U_f^* d}{h^*\sqrt{\phi gh^*}}.
\label{eq:Istar}
\end{equation}
Thus, up to the packing-fraction factor $\phi^{-1/2}$, $I^*$ combines the bulk-scale balance between inertia and gravity, represented by a Froude number $U^*_f/\sqrt{g h^*}$, with the grain-to-layer scale ratio $d/h^*$. It therefore retains the microscopic length scale characteristic of the inertial number and should not be identified with a Froude number alone. This decomposition also shows that the mapping from $a$ to $I^*$ involves
not only the characteristic collapse velocity, but also the relative thickness of the mobilised layer. The inertial-number description therefore incorporates the mass partition associated with the static core through the characteristic thickness $h^*$.

Figure~\ref{fig:Trio_friction}(b) reports the evolution of the experimental values of the effective friction coefficient $\mu_{\rm eff}$ as a function of the macroscopic inertial number $I^*$. 
The coefficient $\mu_{\rm eff}$ increases from approximately 0.5 at small $I^*$ and seems to tend to a plateau value of approximately 0.9 at large $I^*$. Its evolution can be described by the following empirical form

\begin{equation}
{ \mu_{\rm eff}(I^*) = \mu_{\rm -}
+ \frac{\mu_{\rm +}-\mu_{\rm -}}{1+I_0^*/I^*}}.
\label{eq:mu(I)}
\end{equation}

{ This empirical form is inspired by the functional dependence commonly used for the local $\mu(I)$ rheology \citep{Lacaze2009,Andreotti2013} where $\mu_-$, $\mu_+$ and $I_0^*$ are material-dependent parameters. Here, $\mu_-$ and $\mu_+$ denote  the low- and high-$I^*$ asymptotes of the global coefficient $\mu_{\rm eff}$, while $I_0^*$ sets the crossover between them. The best fit of the $\mu_{\rm eff}$ data with \eqref{eq:mu(I)} gives $\mu_{\rm -}\simeq0.39$, $\mu_{\rm +}\simeq0.93$, and $I_0\simeq0.48$. These parameters should not be interpreted as local yield coefficients: the relation $\mu_{\rm eff}(I^*)$ is a macroscopic parametrisation of the collapse-integrated dissipation, not a local constitutive law.}

\section{Comparison with field events}
\label{sec:field_comparison}

{To assess whether the present axisymmetric experiments provide a useful reference for natural landslides, we compare our results with field data reported in the literature. Natural landslides differ substantially from the present laboratory configuration in geometry, scale, material properties and topography. They may also involve polydispersity, cohesion, pore-fluid effects, erosion and entrainment, and therefore cannot be expected to obey strict dynamic similarity with laboratory dry granular columns \citep{Iverson2015}. 
However, we use the field data to compare scaling trends and orders of magnitude as made recently by \cite{CerbusPRL2024}.
Among the landslide types distinguished in the updated Varnes
classification of \citet{Hungr2014}, rock/ice avalanches and dry granular flows are the closest analogues to the present experiments. Large-scale flume experiments provide an intermediate scale between laboratory columns and natural mass movements and further illustrate the influence of pore-pressure evolution and basal conditions \citep{Iverson2010,Ng2026}, while field studies highlight the additional roles of source geometry, topography and effective rheology \citep{Lucas2011,Lucas2014}.}

\citet{Lajeunesse2006Mars} have shown that some Martian landslides obey scaling laws similar to those observed for collapsing cylindrical granular columns in laboratory experiments. A broader compilation by~\cite{Legros2002} gathers numerous subaerial non-volcanic landslides, subaerial volcanic landslides, and landslides on Mars. For these events, the landslide volume $V$, the maximum vertical drop $H_{\max}$, and the maximum horizontal extent $L_{\max}$ are reported. In the present comparison, we retain only those cases for which these three quantities are available, corresponding to 29 non-volcanic landslides, 48 volcanic landslides, and 26 Martian landslides. For a subset of seven well-documented events, summarised in table~\ref{tab:Legros}, the horizontal travel distance of the centre of mass $L$ and its vertical drop $H$ were also given by~\cite{Legros2002}. { These seven events constitute the complete subset of the \citet{Legros2002} compilation for which both components required to estimate the centre-of-mass displacement are available.}

To compare these field events in the framework of through laboratory experiments, we first estimate their mass from the reported volume by assuming a bulk density $\rho = 2.5\times 10^3~\mathrm{kg\,m^{-3}}$, comparable to that of granite and to the density of our glass beads. Although the actual rock types differ between events, their densities are generally of the same order, and the corresponding uncertainty is likely smaller than that associated with poorly constrained topographic effects. We further assume that the entire reported volume is mobilised during the landslide, so the flowing mass is given by $M_f=M_0=\rho V$. The initial potential energy and the energy dissipated during the event are then approximated as
\begin{equation}
E_0 = \frac{1}{2}\,M_0\,g\,H_{\max}, 
\qquad
E_{d \infty} = M_0\,g\,H,
\end{equation}
where $H$ is the vertical drop of the centre of mass. The factor $1/2$ is a crude geometric estimate of the initial centre-of-mass height when only $H_{\max}$ is available; it should not be interpreted as a direct measurement of the potential-energy drop.

For field events, for which no reconstructed deposit geometry is available, we estimate the centre-of-mass displacement from the horizontal and vertical excursions as
\begin{equation}
\Delta_G = \sqrt{L^2+H^2}.
\label{eq:DeltaG_field}
\end{equation}
This estimate should be viewed as the field analogue of the barycentric displacement measured in our laboratory experiments. For terrestrial landslides (subaerial volcanic and non-volcanic), we take $g = 9.81~\mathrm{m\,s^{-2}}$, whereas for Martian landslides we use $g=3.77~\mathrm{m\,s^{-2}}$.

\begin{table}
    \centering
    \begin{tabular}{c||c c c c c c}
         Name & $L_{\max}$ (m) & $H_{\max}$ (m) & $L$ (m) & $H$ (m) & $V$ (m$^3$) & Symbol \\[0.2em]
        \hline\hline\\[-0.5em]
        Blackhawk        & $9\times 10^3$   & $1.17\times 10^3$ & $5.5\times 10^3$  & $1.02\times 10^3$ & $3\times 10^8$    & $\textcolor{ForestGreen}{\squadfill}$  \\
        Elm              & $2\times 10^3$   & $6\times 10^2$    & $9\times 10^2$    & $3.78\times 10^2$ & $1\times 10^7$    & $\textcolor{ForestGreen}{\circletfill}$ \\
        Lastarria        & $6.7\times 10^3$ & $1.01\times 10^3$ & $3.5\times 10^3$  & $6.13\times 10^2$ & $9\times 10^7$    & $\textcolor{BrickRed}{\blacktriangle}$ \\
        Mount St.\ Helens& $2.3\times 10^4$ & $2.07\times 10^3$ & $1\times 10^4$    & $1.85\times 10^3$ & $2.8\times 10^9$  & $\textcolor{BrickRed}{\blacktriangledown}$ \\
        Nevado de Colima & $1.2\times 10^5$ & $4.80\times 10^3$ & $4\times 10^4$    & $3.80\times 10^3$ & $2.7\times 10^{10}$ & $\textcolor{BrickRed}{\bigstar}$ \\
        Shasta           & $4.9\times 10^4$ & $3.43\times 10^3$ & $2.5\times 10^4$  & $2.75\times 10^3$ & $4.5\times 10^{10}$ & $\textcolor{BrickRed}{ \triangleplfill}$ \\
        Sherman          & $6\times 10^3$   & $1.32\times 10^3$ & $3.5\times 10^3$  & $6.65\times 10^2$ & $1\times 10^7$    & $\textcolor{ForestGreen}{\triangleprfill}$ \\
    \end{tabular}
    \caption{Characteristic scales for the seven field events from \citet{Legros2002} for which both horizontal and vertical centre-of-mass displacements are available.}
    \label{tab:Legros}
\end{table}

\begin{figure} [!b]
    \centering
    \begin{subfigure}[b]{0.45\textwidth}
         \centering
         \includegraphics[width=\textwidth]{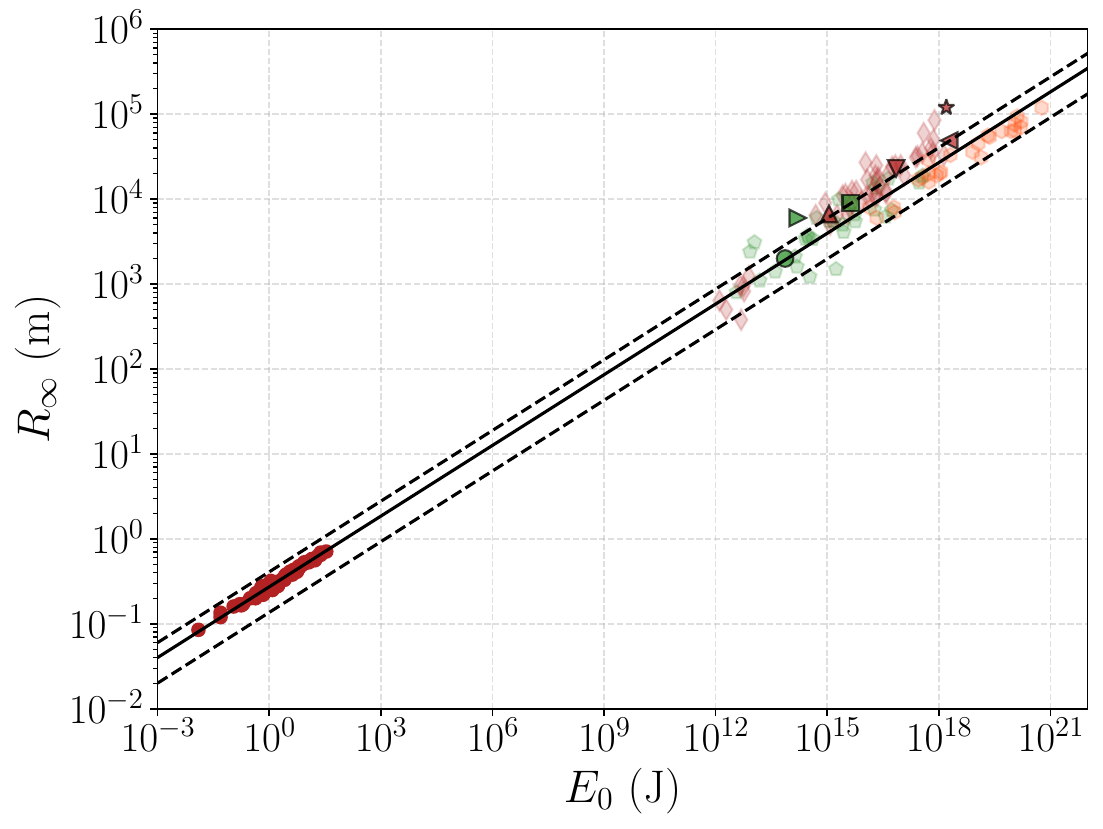}
         \caption{}
         \label{fig:Comparaison_Rf}
     \end{subfigure}
     \hfill
     \begin{subfigure}[b]{0.45\textwidth}
         \centering
         \includegraphics[width=\textwidth]{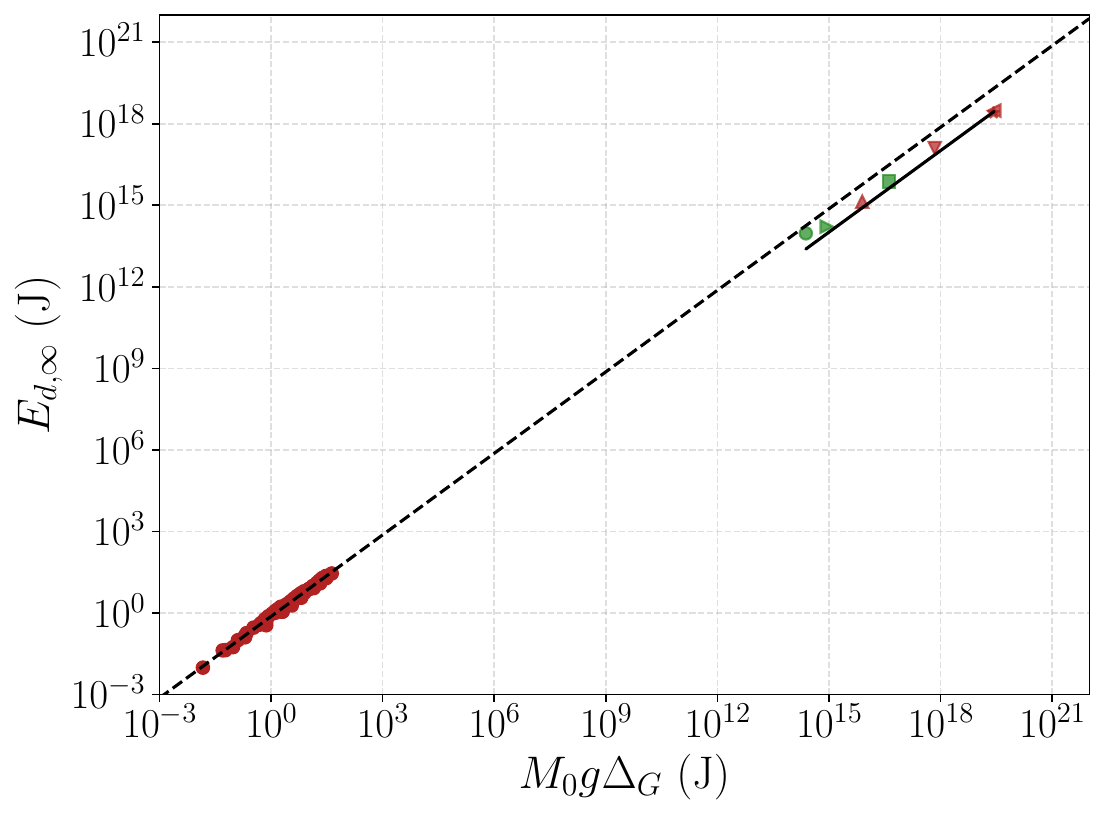}
         \caption{}
         \label{fig:Comparaison_Ediss}
     \end{subfigure}
     \hfill
    \caption{Comparison between laboratory collapses and field landslides. (a) Runout length ($R_\infty$ for laboratory collapses and $L_{\max}$ for field events) as a function of the initial potential energy $E_0$. Laboratory data are shown as red circles (\textcolor{BrickRed}{$\circletfill$}); red, green, and orange symbols correspond to subaerial volcanic, subaerial non-volcanic, and Martian landslides, respectively, as defined by \citet{Legros2002}. The solid black line corresponds to the power-law fit $R_\infty={\beta} E_0^{0.28}$ to the laboratory data, {with $\beta=0.27\,\mathrm{m\,J^{-0.28}}$,} and the dashed lines indicate a $\pm 50\%$ confidence interval. (b) Dissipated energy $E_{\mathrm{diss}}$ as a function of the frictional work $W_0=gM_0\Delta_G$ for laboratory collapses (\textcolor{BrickRed}{$\circletfill$}) { and the seven field events from \citet{Legros2002} for which both horizontal and vertical centre-of-mass displacements are available (symbols defined in Table~\ref{tab:Legros})}. The dashed line (\textcolor{Black}{-{}-}) corresponds to $E_{\mathrm{diss}}=\mu_\textrm{eff} M_0g\Delta_G$ with $\mu_\textrm{eff} =0.74$, and the solid line corresponds to a linear fit of the field data with $\mu_\textrm{eff} =0.1$.}
    \label{fig:Comparaison}
\end{figure}

We first compare the runout length, represented by $R_\infty$ for our laboratory experiments and by $L_{\max}$ for the field landslides, with the initial potential energy $E_0$. Figure~\ref{fig:Comparaison}(a) shows the evolution of both $R_\infty$ (\textcolor{BrickRed}{$\bullet$}) and $L_{\rm max}$ (symbols defined in table~\ref{tab:Legros}) with $E_0$. For the experiments, the best fit
$R_\infty\simeq{\beta}E_0^{0.28}$ from equation~(3.4) is plotted as a solid line together with a $\pm 50\%$ confidence band. Because, for a given meridional geometry, $E_0$ scales linearly with the azimuthal extent $\Delta\theta$, whereas the radial runout is essentially independent of it, the comparison in figure~\ref{fig:Comparaison}(a) between the quarter-cylinder experiments ($\Delta\theta=\pi/2$) and
natural landslides, whose effective spreading angle $\Delta\theta$ varies from one event to another, should primarily be interpreted in terms of scaling trends and orders of magnitude rather than as a quantitative test with a common prefactor. Despite the large diversity of field configurations and the substantial uncertainties in volume estimates and centre-of-mass trajectories, the field data are broadly consistent with an extrapolation of the laboratory trend. If we interpret the runout
length as a characteristic radius, the area covered by the deposit scales as $R_\infty^2 \propto E_0^{0.56}$, which is consistent with the range of exponent
$0.58$--$0.68$ reported by Dade \& Huppert (1998) for the scaling of deposit area with potential energy for natural landslides.

Considering the final potential-energy loss discussed in section~\ref{sec:energetic_dissipation}, laboratory experiments suggest $E_{d \infty} \simeq \mu_{\rm eff}\,gM_0\Delta_{G \infty}$ with $\mu_{\rm eff} \simeq 0.74 \pm 0.11$, so that the dissipated energy is essentially set by a frictional work based on the total mass and the displacement of the barycentre.
Unlike much of the landslide literature, which relates dissipated energy primarily to the overall spreading distance, the present framework suggests
that it is more naturally interpreted in terms of the frictional work associated with the barycentric displacement. For field events, only the seven
landslides listed in table~\ref{tab:Legros} have sufficient information to estimate both $E_{d \infty}$ and the centre-of-mass displacement $\Delta_G$ defined by equation~\eqref{eq:DeltaG_field}. The corresponding
values of $E_{d\infty}$ and $M_0 g \Delta_G$ for laboratory data and field landslides are shown in figure~\ref{fig:Comparaison}(b). The dashed line represents the linear scaling established from the laboratory measurements. When the field data are plotted on the same axes, they also align along an approximately linear trend, but with a smaller coefficient: a linear fit to the field events (solid line) yields
\[
E_{d \infty} \simeq 0.1\,gM_0\Delta_{G,\infty}.
\]
{ Given the small number of events, this comparison should be regarded as a first-order observation rather than as a quantitative validation of the laboratory scaling. Thus, the field data are compatible with an approximately linear dependence
between dissipated energy and the frictional-work estimate, but with an effective proportionality coefficient substantially smaller than in laboratory
slumps.}

{This discrepancy likely reflects the combined effects of large-scale slopes and complex topography, material polydispersity and cohesion, pore-fluid effects, and possible substrate erosion and entrainment, all of which can modify mobility and energy dissipation relative to the present laboratory configuration. Nevertheless, the broadly similar trends observed for laboratory slumps and natural landslides when compared in terms of initial potential energy and frictional work are consistent with the use of the present axisymmetric experiments as a simplified reference for field events.}

In the geophysical literature, frictional work is estimated based on the horizontal movement of the collapse, thereby defining a friction coefficient $\mu=H/L$ as the ratio of the dissipated energy (proportional to $H$) to the frictional work (proportional to $L$) \citep{DadeHuppert1998,StaronLajeunesse2009}. In this study, we consider the global displacement $\Delta_G = \sqrt{L^2+H^2}$ which combines both the horizontal and vertical displacements during the collapse. However, as geophysical collapses are analogous to column collapses with a low aspect ratio \citep{Lajeunesse2006Mars} where the displacements are mainly horizontal, these two approaches are to some extent equivalent. { The barycentric energy balance is the element of the present framework that can be compared most directly with field data when the mobilised mass and centre-of-mass displacement can be estimated. By contrast, applying the specific static-core closure to a natural event would require independent characterisation of the actual mobilised region, topography and basal conditions.}

\section{Conclusion}
\label{sec:conclusion}

We have experimentally investigated the collapse of an axisymmetric granular column of aspect ratio $a=H_0/R_0$ and mass $M_0$, focusing on the geometry of the final deposit, the partitioning of mass between mobile and static grains, and the motion of the barycentre of an elementary azimuthal slice, considering either the entire granular mass or the flowing grains alone. Varying the initial aspect ratio over almost two decades, three regimes of collapse {already reported in the literature} are observed: a truncated-cone regime at low aspect ratio, a conical regime at intermediate aspect ratio, and a granular-jet-like regime at large aspect ratio. In regimes~I and II, the normalised final height follows the following  piecewise laws already reported in the litterature with $H_\infty/R_0 = a$ at small aspect ratio and a plateau $H_\infty/R_0 = C \simeq 0.79$ above a critical aspect ratio $a_c$. In regime II, the final radius is well described by the scaling law $R_\infty/R_0 = D\,a^{1/2}$ with $D \simeq 2.4$ as a function of the initial aspect ratio $a$, and also close to the scaling $R_\infty\propto E_0^{1/4}$ as a function of the initial potential energy. We show that these scaling laws can be rationalised using a simplified geometric picture in which a static inner region remains immobile throughout the collapse: a static truncated cone in regime I and a static full cone in regime II with an approximately constant slope angle, $\theta_2 \simeq (37 \pm 4 )^\circ$, that appears approximately independent of the initial aspect ratio. Within this framework, the scaling exponents and the order of magnitude of their prefactors can be predicted analytically. The height plateau is directly related to the static-interface angle $\theta_2$, with $C=H_\infty/R_0 \simeq \tan\theta_2$, while the transition between regimes I and II occurs at $a_c \simeq \tan\theta_2$. Volume conservation between the initial cylindrical column and the final conical deposit then yields the square-root runout law $R_\infty/R_0=D\,a^{1/2}$ in regime II, with the model prediction $D_{\mathrm{model}}\simeq(3/\tan\theta_2)^{1/2}\simeq2.1$, reasonably close to the measured value $D\simeq2.4$. In regime I, similar volume conservation leads to another predicted scaling that deviates slightly from the scaling of regime II. This highlights that global deposit dimensions in an axisymmetric configuration can be captured quantitatively through simple geometric considerations based on the static core, without resorting to detailed modelling of the transient flow. The model also yields analytical expressions for the flowing- and exiting-mass fractions as functions of the initial aspect ratio. These expressions capture the measurements well and explain why, at sufficiently large aspect ratios, the exiting mass provides a good estimate of the flowing mass. Tracking the barycentre displacement $\Delta_G$ of an elementary azimuthal slice, for either the entire granular mass or the flowing grains alone, provides an estimate of the dissipated energy and hence of a global effective friction coefficient $\mu_{\mathrm{eff}}$. This coefficient increases from approximately $0.5$ at low $a$ to about $0.9$ at large $a$, a trend broadly reproduced by the model. {
A macroscopic inertial number $I^*$, built from characteristic estimates of the shear rate and mobilised-layer thickness, provides an empirical $\mu_{\rm eff}(I^*)$ parametrisation of this evolution, inspired by the functional form of local $\mu(I)$ rheology.}%

Finally, a comparison with natural landslides, using compiled field data for terrestrial and Martian events, indicates that the laboratory scalings provide a useful reference despite the complexity of natural topography and material properties. Field runout data are broadly compatible with an extrapolation of the sublinear laboratory trend. For the subset of seven events for which a barycentric displacement can be estimated, the dissipated energy also varies approximately linearly with $M_0 g\Delta_G$, although with a substantially smaller proportionality coefficient than in the laboratory. Altogether, the present study provides a coherent framework linking deposit geometry, mass partition, and energy dissipation in axisymmetric granular slumps. It also provides geometric and energetic constraints that may help inform reduced or geophysical models of dry granular failures.

{All the modelling developed here relies on the approximately constant slope of the static interface observed in regimes I and II of the collapse of axisymmetric columns, but it leaves several open questions. A first question concerns regime III. Our preliminary observations reveal that the static zone still exists, but with a slope angle $\theta_2$ that decreases with increasing aspect ratio $a$. However, a dedicated study is necessary to investigate this striking regime in detail, under conditions of sufficiently large $R_0/d$ to avoid finite-size effects.

Another important question concerns the planar (two-dimensional) configuration, where a stagnant zone also exists during the collapse of rectangular columns \citep{Staron2005,Lacaze2008}, but with a non-constant slope $\theta_2$ in regime II, leading to a normalised final height that is not constant but increases approximately as $a^{1/3}$ \citep{Lajeunesse2005Slumping,Sarlin2021PRE}. Thus, our modelling based on a constant slope $\theta_2$ in the axisymmetric geometry cannot be transferred directly to the planar geometry, which deserves a dedicated study. More generally, it would also be interesting to investigate how the static zone is modified in non-axisymmetric three-dimensional configurations, such as rectangular columns unconfined by lateral walls, which lead to intriguing non-axisymmetric deposits as reported by \cite{Man2022}.

The present energy-based description therefore complements transient approaches by providing a direct estimate of the total energy loss from the initial and final static states. To investigate further the dissipation processes, measurements from Particle Image Velocimetry (PIV) could also be useful, e.g., for quantifying the instantaneous kinetic energy and thus the instantaneous dissipated energy.

Finally, the present framework developed for dry granular axisymmetric collapse provides estimates of the mobilised mass and of its geometrical and energetic characteristics, which may constitute useful inputs for future models of impulse-wave generated by landslides \citep{Sarlin2021JFM,Sarlin2022PRF}. 
}

{ }

\bigbreak
{
\backsection[Supplementary material]{Supplementary movies show representative experiments in regimes~I, II and III, including the gate-opening stage, the subsequent collapse and the final deposition. The data underlying the figures are also provided as supplementary material.}}

\backsection[Acknowledgments]{The authors thank Wladimir Sarlin and Jean-Gabriel Pichon for their contribution to the initial development of the experimental set-up and for carrying out preliminary experiments. The authors also thank Johannes Amarni, Alban Aubertin, Lionel Auffray, Christophe Manquest, and Rafael Pidoux for their technical support in building and maintaining the experimental set-up.}

\backsection[Funding]{This work has received financial support from the French Research Agency through the Slide2Wave project ANR-23-CE30-0052-01.}

\backsection[Declaration of interests]{The authors report no conflict of interest.}

\backsection[Author ORCIDs]{Igor Le Pivert-{}-Jolivet, \url{https://orcid.org/0009-0001-2531-5098};
Cyprien Morize, \url{https://orcid.org/0000-0002-6966-648X};
Alban Sauret, \url{https://orcid.org/0000-0001-7874-5983};
Philippe Gondret, \url{https://orcid.org/0000-0002-7184-9429}.}

\appendix
\section{Detailed calculations of the masses $M_f$ and $M_{e\infty}$ and of the positions of the slice barycentres $G_\infty$ and $G_{f\infty}$}
\label{appendix:masses}

In this appendix, we detail the geometric calculations leading to the analytical expressions of the final exiting mass $M_{e \infty}$, of the flowing mass $M_{f}$ and of the final positions $G_\infty$ and $G_{f\infty}$ of the ``slice'' barycentres $G$ and $G_f$ corresponding to all the grains or only the flowing grains used in \S\ref{sec:mass_energy}. Throughout, we assume a homogeneous packing fraction $\phi$ and negligible dilation, so that masses $M$ are proportional to volumes $V$: $M=\rho\phi V$. {We also consider the axisymmetry so that all volumes, masses and energies are proportional to the angular extent $\Delta\theta$. In the following, the calculations are performed for a cylindrical sector of arbitrary angular extent $\Delta\theta$, with initial volume $V_0=\Delta\theta R_0^2H_0/2$. For the quarter-cylinder geometry used in the present experiments, $\Delta\theta=\pi/2$, while $\Delta\theta=\pi$ and $\Delta\theta=2\pi$ correspond to a half-cylinder and a full cylinder, respectively.
It is worth noting that this angular extent cancels in the mass fractions $M_f/M_0$ and $M_{e\infty}/M_0$. The barycentres considered here are those of an elementary azimuthal slice $d \theta$, for which the common factor ${\rm d}\theta$ likewise cancels from the ratios defining their radial and vertical coordinates. The ``slice barycentre'' is considered rather than the global barycentre corresponding to all the angular extent $\Delta \theta$, as the radial displacement of the global barycentre would not reflect the radial spreading well. For example, for a full cylinder of angular extent $\Delta\theta=2\pi$, the global barycentre would remain on the symmetry axis during the collapse and would therefore have zero radial displacement, despite the radial spreading of the column.}

\subsection{Regime I: truncated-cone deposits ($a<a_c$)}
\label{appendix:regimeI}

In regime I, the deposit is idealised as a truncated cone of height $H_0$, with inner truncation radius $R_c$ and outer radius $R_\infty^{\mathrm{I}}$ (figure~\ref{fig:Modele}a). Its free surface $h^{\mathrm{I}}(r)$ is then given by

\begin{equation}
h^{\mathrm{I}}(r,t_\infty)=
\begin{cases}
H_0 & \text{for} \, \, r\leq R_c, \\[0.2em]
H_0\,\frac{R_\infty^{\mathrm{I}}-r}{R_\infty^{\mathrm{I}}-R_c} & \text{for} \, \, R_c\leq r\leq R_\infty^{\mathrm{I}},
\end{cases}
\label{eq:app_h_regI}
\end{equation}

\noindent with the truncation radius $R_c = R_0(1 - a/a_c)$ as given by  \eqref{eq:Rc} and \eqref{eq:transi}. The static region corresponds to a truncated cone of height $H_0$ with inner truncated radius $R_c$ and outer radius $R_0$ where the interface is given by
\begin{equation}
h_s^{\mathrm{I}}(r)=
\begin{cases}
H_0, & \text{for} \, \, r\leq R_c, \\[0.2em]
H_0\,\dfrac{R_0-r}{R_0-R_c}, & \text{for} \, \, R_c \leq r\leq R_0.
\end{cases}
\label{eq:app_hf_regI}
\end{equation}

\subsubsection{Flowing mass $M_f^{\mathrm{I}}$}
\label{appendix:regimeI_flowing}

The volume of the flowing grains corresponds to the annulus $R_c \leq r\leq R_0$ above the static part $z=h_s^{\rm I}(r)$ in the initial column of height $H_0$ so that the flowing volume is given by
\begin{equation}
V_f^{\mathrm{I}}
=\int_{0}^{\Delta \theta}\int_{R_c}^{R_0}\bigl(H_0-h_s^{\rm I}(r)\bigr)\,r\,\mathrm{d}r\,\mathrm{d}\theta =\Delta \theta \, H_0 \,\frac{\displaystyle R_0^3/3 -R_cR_0^2/2+R_c^3/6}{R_0-R_c},
\label{eq:app_VfI}
\end{equation}
\noindent {for any angular extent $\Delta \theta$ from 0 to $2 \pi$}.
Using $R_c = R_0(1 - a/a_c)$ from \eqref{eq:Rc} and \eqref{eq:transi} and normalising $V_f^{\mathrm{I}}$ by $V_0$ yields the fraction of flowing mass
\begin{equation}
\frac{M_f^{\mathrm{I}}}{M_0} 
=\frac{V_f^{\mathrm{I}}}{V_0}
= \frac{a}{a_c}\left(1-\frac{a}{3 a_c}\right).
\label{eq:app_MfI_Rc}
\end{equation}
corresponding to \eqref{eq:MfI_a}.

\subsubsection{Exiting mass $M_{e\infty}^{\mathrm{I}}$}
\label{appendix:regimeI_exiting}

The exiting mass corresponds to the part of the deposit lying outside the initial footprint, \ie\ in the region $r>R_0$. Using \eqref{eq:app_h_regI}, this exiting volume is
\begin{equation}
V_{e\infty}^{\mathrm{I}}
=\int_{0}^{\Delta \theta}\int_{R_0}^{R_\infty^{\mathrm{I}}} h^{\rm I}(r,t_\infty)\,r\,\mathrm{d}r\,\mathrm{d}\theta
=\frac{\Delta \theta H_0}{6}\,\frac{\displaystyle {R_\infty^{\mathrm{I}}}^3- 3 R_\infty^{\mathrm{I}}R_0^2+2 R_0^3}{R_\infty^{\mathrm{I}}-R_c}.
\label{eq:app_VsI_def}
\end{equation}
Normalising $V_{e\infty}^{\mathrm{I}}$ by the initial volume $V_0 = \Delta \theta R_0^2 H_0/2$ leads first to
\begin{equation}
\frac{M_{e\infty}^{\mathrm{I}}}{M_0}
=\frac{V_{e\infty}^{\mathrm{I}}}{V_0}
=\frac{(R_\infty^{\mathrm{I}}/R_0-1)^{2}(R_\infty^{\mathrm{I}}/R_0+2)}{3\,[R_\infty^{\mathrm{I}}/R_0-R_c/R_0]}.
\label{eq:app_MsI_X}
\end{equation}
Using \eqref{eq:Rc} with \eqref{eq:transi} for $R_c/R_0$ and \eqref{eq:Rf_regimeI} for $R_\infty^{\rm I}/R_0$, \eqref{eq:app_MsI_X} leads finally to
\begin{equation}
\frac{M_{e \infty}^{\rm I}}{M_0}
= \frac{2}{3}
\frac{
 \left(3-a/a_c\right) (a/a_c)^2
}{\sqrt{12-3\left( 1-a/a_c\right)^2}  -3\left( 1-a/a_c\right)}  \quad  \text{for } a \leqslant a_c.
\end{equation}

\subsection{Regime II: conical deposits ($a_c<a\lesssim 12$)}
\label{appendix:regimeII}

In regime~II, the final deposit is idealised as a cone of height $H_\infty$ and radius $R_\infty$ (figure~\ref{fig:Modele}(a)), with the free surface 
\begin{equation}
h^{\mathrm{II}}(r,t_\infty)=H_\infty^{\rm II}\left(1-\frac{r}{R_\infty^{\mathrm{II}}}\right)
\qquad \text{for} \, \, 0\le r\le R_\infty^{\mathrm{II}},
\label{eq:app_h_regII_free}
\end{equation}
while the static region is an inner cone of radius $R_0$ and height $H_\infty$ (figure~\ref{fig:Modele}a) with the interface 
\begin{equation}
h_s^{\mathrm{II}}(r)=H_\infty^{\rm II}\left(1-\frac{r}{R_0}\right)
\qquad \text{for} \, \, 0\le r\le R_0,
\label{eq:app_h_regII_static}
\end{equation}

\subsubsection{Flowing mass $M_f^{\mathrm{II}}$}
\label{appendix:regimeII_flowing}

As the static volume in regime II corresponds to a conical part of base radius $R_0$ and height $H_\infty$ of volume $\Delta \theta \, R_0^2 \, H_\infty/6$,
the flowing mass $M_f^{\mathrm{II}}$ in regime II corresponds simply to the difference between the mass of the initial cylindrical part of volume $V_0=\Delta \theta\, R_0^2 \, H_0/2$ and that of the static conical part, so that
\begin{equation}
\frac{M_f^{\mathrm{II}}}{M_0}=\frac{V_f^{\mathrm{II}}}{V_0}
=1-\frac{H_\infty^{\rm II}}{3H_0}.
\label{eq:app_MfII_Hf}
\end{equation}
Using \eqref{eq:Hf_regimeII} with $C \simeq a_c$ for $H_\infty^{\rm II}/H_0$, \eqref{eq:app_MfII_Hf} yields simply to
\begin{equation}
\frac{M_{f}^{\mathrm{II}}}{M_0} \simeq 1-\frac{a_c}{3a}  \quad  \text{for } a \geqslant a_c.
\end{equation}

\subsubsection{Exiting mass $M_{e\infty}^{\mathrm{II}}$}
\label{appendix:regimeII_exiting}

The volume $V_{\mathrm{in}}^{\mathrm{II}}$ of grains that remain within the initial footprint corresponds to the intersection of the final cone with the cylinder $r\le R_0$. Using \eqref{eq:app_h_regII_free}, the corresponding volume is
\begin{equation}
V_{\mathrm{in}\infty}^{\mathrm{II}}
=\int_{0}^{\Delta \theta}\int_{0}^{R_0} h^{\mathrm{II}}(r,t_\infty)\,r\,\mathrm{d}r\,\mathrm{d}\theta \nonumber\\
=\Delta \theta\, H_\infty^{\rm II}\left(\frac{R_0^2}{2}-\frac{R_0^3}{3R_\infty^{\rm II}}\right),
\label{eq:app_VinII}
\end{equation}
and the exiting mass $M_{e\infty}^{\mathrm{II}}$ therefore reduces to
\begin{equation}
\frac{M_{e\infty}^{\mathrm{II}}}{M_0}
=1-\frac{V_{\mathrm{in\infty}}^{\mathrm{II}}}{V_0}
=1-\frac{H_\infty^{\rm II}}{H_0}\left(1-\frac{2R_0}{3R_\infty^{\rm II}}\right).
\label{eq:app_MsII_general}
\end{equation}
Using \eqref{eq:Hf_regimeII} with $C \simeq a_c$ for $H_\infty^{\rm II}/H_0$ together with \eqref{eq:Rf_regimeII} for $R_\infty^{\rm II}/R_0$, \eqref{eq:app_MsII_general} finally leads to 
\begin{equation}
\frac{M_{e \infty}^{\mathrm{II}}}{M_0}
\simeq 1-\frac{a_c}{a}+\frac{2}{3\sqrt{3}}\,\left(\frac{a_c}{a}\right)^{3/2}  \quad  \text{for } a \geqslant a_c.
\end{equation}

\subsection{``Slice'' barycentres}
\label{app:barycentre_3D}

\subsubsection{``Slice'' barycentre $G$ of all grains}
\label{app:barycentre_allgrains}

{ For an axisymmetric deposit described by the instantaneous profile $h(r,t)$, the volume element is ${\rm d}V=r\,{\rm d}r\,{\rm d}\theta\,{\rm d}z$. Assuming a homogeneous packing fraction within the slice, the vertical and radial coordinates of the ``slice'' barycentre are
}
\begin{subequations}
\begin{align}
 z_G(t) &=
 \frac{\displaystyle \int_{r=0}^{R(t)} \int_{z=0}^{h(r,t)} z\,{\rm d}z\,r\,{\rm d}r\,{\rm d}\theta}
 {\displaystyle \int_{r=0}^{R(t)} \int_{z=0}^{h(r,t)} {\rm d}z\,r\,{\rm d}r\,{\rm d}\theta}
 =
 \frac{\displaystyle \int_0^{R(t)} h^2(r,t)\,r\,{\rm d}r}
 {\displaystyle 2\int_0^{R(t)} h(r,t)\,r\,{\rm d}r},
 \label{eq:def_zG}
 \\
 r_G(t) &=
 \frac{\displaystyle \int_{r=0}^{R(t)} \int_{z=0}^{h(r,t)} r\,{\rm d}z\,r\,{\rm d}r\,{\rm d}\theta}
 {\displaystyle \int_{r=0}^{R(t)} \int_{z=0}^{h(r,t)} {\rm d}z\,r\,{\rm d}r\,{\rm d}\theta}
 =
 \frac{\displaystyle \int_0^{R(t)} h(r,t)\,r^2\,{\rm d}r}
 {\displaystyle \int_0^{R(t)} h(r,t)\,r\,{\rm d}r}.
 \label{eq:def_rG}
\end{align}
\end{subequations}

In regime~I, using the piecewise profile $h^{\mathrm{I}}(r,t_\infty)$ given by \eqref{eq:app_h_regI}, the volume integral and the two first moments corresponding to the position $(r_{G\infty}^{\mathrm{I}}, z_{G\infty}^{\mathrm{I}})$ of the barycentre $G_\infty$ for the final deposit are
\begin{subequations}
\begin{align}
\int_0^{R_\infty^{\mathrm{I}}} h^{\mathrm{I}}(r,t_\infty)\,r\,{\rm d}r &= \frac{H_0}{6} \left({R_\infty^{\mathrm{I}}}^2 + R_\infty^{\mathrm{I}}R_c + {R_c}^2 \right),\\
\int_0^{R_\infty^{\mathrm{I}}} h^{\mathrm{I}}(r,t_\infty)^2\,r\,{\rm d}r &= \frac{H_0^2}{12} \left({R_\infty^{\mathrm{I}}}^2 + 2R_\infty^{\mathrm{I}}R_c + 3{R_c}^2 \right),\\
\int_0^{R_\infty^{\mathrm{I}}} h^{\mathrm{I}}(r,t_\infty)\,r^2\,{\rm d}r &= \frac{H_0}{12} \left( R_\infty^{\mathrm{I}}+R_c \right) \left({R_\infty^{\mathrm{I}}}^2+{R_c}^2 \right).
\end{align}
\end{subequations}

It follows that
\begin{equation}
z_{G\infty}^{\mathrm{I}} = \frac{H_0}{4} \frac{
{R_\infty^{\mathrm{I}}}^2 + 2R_\infty^{\mathrm{I}}R_c + 3{R_c}^2}{{R_\infty^{\mathrm{I}}}^2 +
R_\infty^{\mathrm{I}}R_c + {R_c}^2},
\qquad
r_{G\infty}^{\mathrm{I}} = \frac{ \left(R_\infty^{\mathrm{I}}+R_c\right)
\left({R_\infty^{\mathrm{I}}}^2+{R_c}^2
\right)}{2\left( {R_\infty^{\mathrm{I}}}^2
+ R_\infty^{\mathrm{I}}R_c + {R_c}^2 \right) }.
\label{eq:calculbaryI}
\end{equation}
Using \eqref{eq:Rc}-\eqref{eq:transi} for $R_c$ and \eqref{eq:Rf_regimeI} for $R_\infty^{\rm I}$, \eqref{eq:calculbaryI} leads to the normalised final positions
\begin{equation}
\begin{aligned}
\frac{z_{G\infty}^{\mathrm{I}}}{H_0} &= \frac{ 6+3\left(1-a/a_c\right)^2+\left(1-a/a_c\right) \sqrt{12-3\left(1-a/a_c\right)^2}}{24} \quad  \text{for } a \leqslant a_c,
\\
\frac{r_{G\infty}^{\mathrm{I}}}{R_0} &= \frac{ 3\sqrt{12-3\left(1-a/a_c\right)^2}-3\left(1-a/a_c\right) +2\left(1-a/a_c\right)^3 }{12} \quad  \text{for } a \leqslant a_c.
\end{aligned}
\end{equation}

In regime~II, the volume integral and the two first moments corresponding to the position {$(r^{II}_{G\infty},z^{II}_{G\infty})$} of the barycentre $G_\infty$ for the final deposit of profile $h^{\mathrm{II}}(r,t_\infty)$ given by \eqref{eq:app_h_regII_free} are
\begin{subequations}
\begin{align}
\int_0^{R_\infty^{\mathrm{II}}} r\,h^{\mathrm{II}}(r,t_\infty)\,{\rm d}r &= \frac{H_\infty^{\mathrm{II}}{R_\infty^{\mathrm{II}}}^2}{6},\\
\int_0^{R_\infty^{\mathrm{II}}} r\,h^{\mathrm{II}}(r,t_\infty)^2\,{\rm d}r &= \frac{{H_\infty^{\mathrm{II}}}^2{R_\infty^{\mathrm{II}}}^2}{12},\\
\int_0^{R_\infty^{\mathrm{II}}} r^2\,h^{\mathrm{II}}(r,t_\infty)\,{\rm d}r &=
\frac{H_\infty^{\mathrm{II}}{R_\infty^{\mathrm{II}}}^3}{12}.
\end{align}
\end{subequations}
Thus, the barycentre of any elementary angular slice of the conical deposit of regime II is located at
\begin{equation}
z_{G_\infty}^{\mathrm{II}}=\frac{H_\infty^{\mathrm{II}}}{4}
\quad {\rm and} \quad
r_{G_\infty}^{\mathrm{II}}=\frac{R_\infty^{\mathrm{II}}}{2}.
\label{eq:calculbaryII}
\end{equation}
Using \eqref{eq:Hf_regimeII} with $C \simeq a_c$ for $H_\infty^{\rm II}$ and \eqref{eq:Rf_regimeII} for $R_\infty^{\rm II}$, \eqref{eq:calculbaryII} leads to the normalised final positions
\begin{equation}
\frac{z_{G\infty}^{\mathrm{II}}}{H_0} \simeq \frac{a_c}{4a} 
\quad
\text{and}
\quad
 \frac{r_{G\infty}^{\mathrm{II}}}{R_0} = \frac{1}{2}\left(\frac{3a}{a_c}\right)^{1/2} \quad  \text{for } a \geqslant a_c.
\end{equation}

\subsubsection{``Slice'' barycentre $G_f$ of flowing grains}
\label{app:barycentre_flowinggrains}
From the instantaneous measured interface $h(r,t)$ and static interface $h_s(r)$ can be extracted the instantaneous barycentre $G_f(t)=(r_{G_f}(t),z_{G_f}(t))$ of the flowing grains contained in any slice of elementary azimuthal extension. Considering that the packing fraction $\phi$ is homogeneous within the slice, the position of this slice barycentre is given by
\begin{subequations}
\begin{align}
  z_{G_f}(t) = \frac{\displaystyle \int_{r=0}^{R(t)} \int_{z=h_s(r)}^{h(r,t)} z\,{\rm d}z\,r\,{\rm d}r \,{\rm d}\theta}{\displaystyle \int_{r=0}^{R(t)} \int_{z=h_s(r)}^{h(r,t)} {\rm d}z\,r\,{\rm d}r \,{\rm d}\theta}
 = \frac{\displaystyle \int_0^{R(t)} [h^2(r,t) - h_s^2(r)]\,r\,{\rm d}r }{\displaystyle 2\int_0^{R(t)} [h(r,t)-h_s(r)]\,r\,{\rm d}r},
\\
r_{G_f}(t) = \frac{\displaystyle \int_{r=0}^{R(t)} \int_{z=h_s(r)}^{h(r,t)} r\,{\rm d}z\,r\,{\rm d}r \,{\rm d}\theta}{\displaystyle \int_{r=0}^{R(t)} \int_{z=h_s(r)}^{h(r,t)} {\rm d}z\,r\,{\rm d}r \,{\rm d}\theta}
= \frac{\displaystyle \int_0^{R(t)} [h(r,t)-h_s(r)]\,r^2\,{\rm d}r}{\displaystyle \int_0^{R(t)} [h(r,t)-h_s(r)]\,r\,{\rm d}r}. 
\end{align} 
\end{subequations}

\section{Characteristic flowing thickness $h^*$}
\label{app:hstar}

\begin{figure}
    \centering
    \begin{subfigure}[b]{0.45\textwidth}
         \centering
         \includegraphics[width=\textwidth]{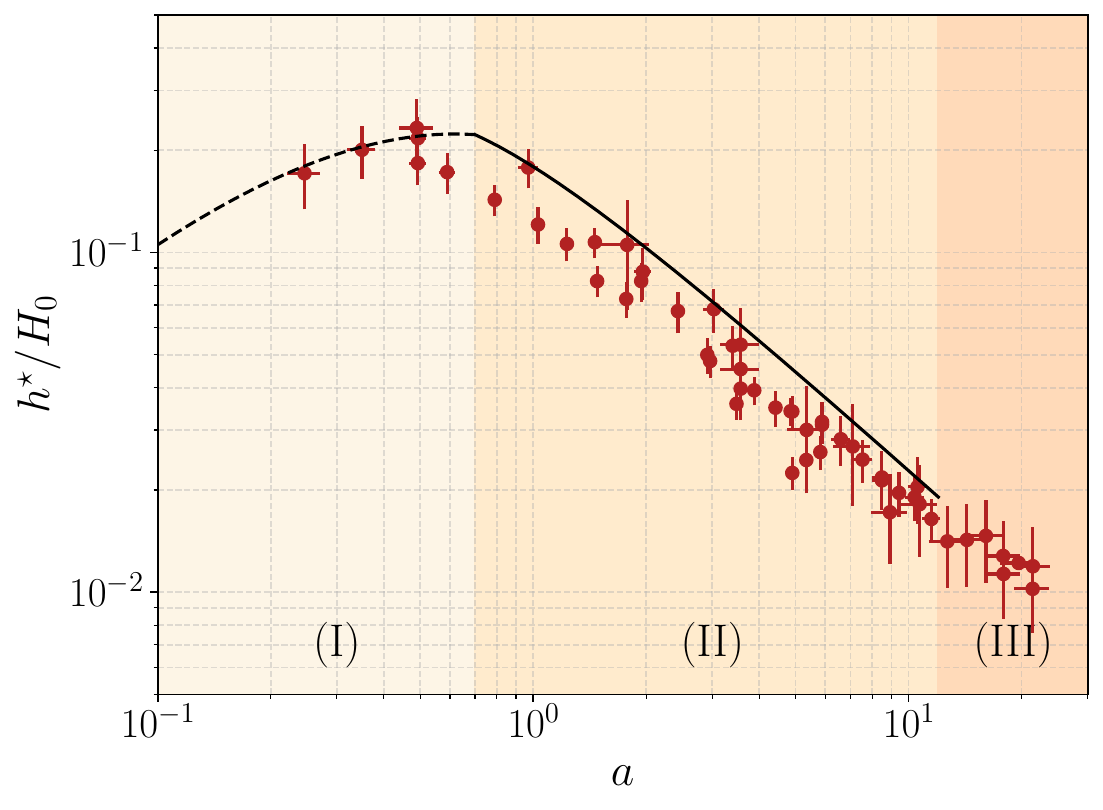}
         \label{fig:hstar_H0}
     \end{subfigure}
     \hfill
     \begin{subfigure}[b]{0.45\textwidth}
         \centering
         \includegraphics[width=\textwidth]{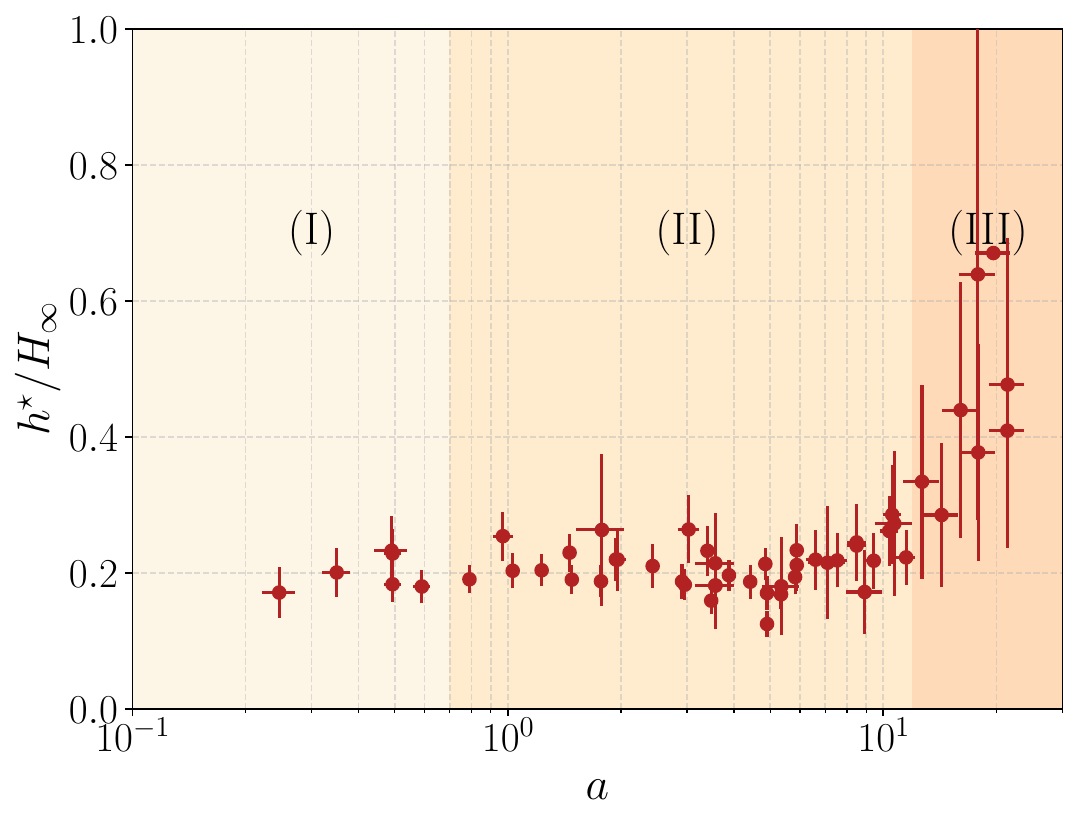}
         \label{fig:hstar_Hf}
     \end{subfigure}
     \hfill
    \caption{Characteristic thickness $h^*$ normalised either by (a) the initial height $H_0$ of the column or (b) by the final height $H_\infty$ of the deposit as a function of the aspect ratio $a$. The data points are the experimental measurements, whereas the dashed and solid lines are the model predictions given by \eqref{eq:h*regimeI} and \eqref{eq:h*regimeII} with $a_c=0.7$.}
    \label{fig:hstar}
\end{figure}

Considering the flowing volume $V_f/V_0 = M_f/M_0$ with $V_0=\Delta \theta\, R_0^2 \, H_0/2$, and the characteristic area $A^*=\Delta \theta\, R_\infty^2 /2$ of the deposit for any angular extent $\Delta \theta$ of the initial cylindrical column ($\Delta \theta = \pi/2$ for our particular quarter-column), the characteristic flowing thickness $h^* = V_f/A^*$ introduced in \S \ref{sec:muI} is simply given by
\begin{equation}
\frac{h^*}{H_0} = \left(\frac{R_0}{R_\infty}\right)^2 \frac{M_f}{M_0}.
\label{eq:hstar}
\end{equation}
The characteristic shear thickness $h^*$, normalised either by the initial height $H_0$ of the column or by the final height $H_\infty$ of the deposit, is shown as a function of the aspect ratio in figure~\ref{fig:hstar}(a) and \ref{fig:hstar}(b), respectively. The measurements of $h^*/H_0$ show a plateau of value 0.2 in regime I followed by a decrease in regime II with a slope of $-1$ in the log--log plot in figure~\ref{fig:hstar}(a). The characteristic shear thickness $h^*$ scales thus approximately as $0.2 \, H_0$ in regime I and as $0.2  \, R_0$ in regime II. The measurements of $h^*/H_\infty$ displayed in figure~\ref{fig:hstar}(b) show a nice plateau over both regimes I and II: $h^* \simeq 0.2  \, H_\infty$. This is no longer the case in regime III where the height $H_\infty$ at the axis of symmetry is smaller than the height of the bump off the axis of symmetry. In regime III, the characteristic flowing thickness should be instead the bump height.

The prediction of the model for $h^*/H_0$ can be obtained by considering \eqref{eq:hstar} together with \eqref{eq:Rf_regimeI} and \eqref{eq:MfI_a} in regime I or with \eqref{eq:Rf_regimeII} and \eqref{eq:MfII} in regime II, which leads to
\begin{equation}
\frac{h^{*{\rm I}}}{H_0} = \frac{4 (a/a_c) \left(1-a/(3a_c)\right)}{\left[\sqrt{12-3\left(1-a/a_c\right)^2}-\left(1-a/a_c\right)\right]^2}
\qquad \text{for } a \le a_c.
\label{eq:h*regimeI}
\end{equation}
\begin{equation}
\frac{h^{* \rm II}}{H_0} =\frac{a_c}{3a}\left( 1 - \frac{a_c}{3a}\right) \quad  \text{for } a \geqslant a_c.
\label{eq:h*regimeII}
\end{equation}

The model equations \eqref{eq:h*regimeI} and \eqref{eq:h*regimeII} displayed in dashed and solid lines in figure \ref{fig:hstar}(a) for $a_c = 0.7$ are quite close to the experimental data. In particular, the value $h^*/H_0 = 2/9 \simeq 0.22$ predicted by the model at $a=a_c$ is close to the observed plateau 0.2 of regime I and the predicted scaling $h^{* \rm II}/H_0 \simeq a_c/3a \simeq 0.23/a$ for $a \gg a_c$ is also very close to the observed experimental decrease of $h^*/H_0$ with $a$ in regime II, thus corresponding to $h^{* \rm II} \simeq 0.23 \, R_0$. However, the prediction of the model in regime I does not correspond to a plateau but to the linear scaling $h^{*\mathrm{I}}/H_0 \simeq a/a_c \simeq 1.4\,a$.
\bibliographystyle{jfm}
\bibliography{jfm}

\end{document}